\documentclass{aa}

\usepackage{float} % control float objects location

\usepackage{textcomp}
\usepackage[utf8]{inputenc}

\usepackage{amsmath, amssymb, amsfonts, amscd}
\usepackage{mathtools} % required by empheq
\usepackage{array} % array environment (in mathmode)
\usepackage{cases} % a math environnement for cases
\usepackage{empheq} % a math environnement allowing letflbrace (ex.)
\usepackage{multirow} % split array cells into mutiple lines or columns
\usepackage{txfonts} % math symbols, as bold Times (next amsmath)

\usepackage{graphicx} % to include graphics
\usepackage{wrapfig} % allows text to be wrapped around floating objects

\usepackage[usenames,dvipsnames]{color} % to color references
\usepackage{ulem} % provides various types of underlining
\usepackage[colorlinks=true,urlcolor=blue,citecolor=blue]{hyperref} % allows references
\usepackage{natbib}
\usepackage[toc,page,titletoc]{appendix}
\usepackage{colortbl}

\begin{document}

%==TITLE==%=========%=========%=========%=========%=========%=========%=
\title{\Large Characterizing the propagation of gravity waves \\ in 3D nonlinear simulations of solar-like stars}

%=AUTHORS=%=========%=========%=========%=========%=========%=========%=
\author{L. Alvan \inst{1}, A. Strugarek \inst{2,1}, A.S. Brun \inst{1}, S. Mathis \inst{1}, R.A. Garcia \inst{1}}
\offprints{L. Alvan}

%==INSTITUTES==%=========%=========%=========%=========%=========%=========%=
\institute{
\inst{1} Laboratoire AIM Paris-Saclay, CEA/DSM - CNRS - Universit\'e Paris Diderot, IRFU/SAp Centre de Saclay, F-91191 Gif-sur-Yvette, France \\
\inst{2} Département de physique, Université de Montréal, C.P. 6128 Succ. Centre-Ville, Montréal, QC H3C-3J7, Canada
\\{}
 \email{lucie.alvan@cea.fr, strugarek@astro.umontreal.ca, sacha.brun@cea.fr, stephane.mathis@cea.fr, rafael.garcia@cea.fr} 
}

%==DATES==%=========%=========%=========%=========%=========%=========%=
%\date{Received ; accepted}
\date{Received : 02/04/2015 ; accepted : 04/06/2015}

%=ABSTRACT==%=========%=========%=========%=========%=========%=========%=
\abstract
% Context
{The revolution of helio- and asteroseismology provides access to {the detailed properties of} stellar interiors by studying the star's
  oscillation modes. Among them, {gravity (g)} modes are formed by constructive interferences between progressive internal gravity waves (IGWs),
  propagating in stellar radiative zones. Our new 3D nonlinear simulations of the interior of a solar-like star allows us to study the excitation,
  propagation, and dissipation of these waves. 
}
% Aims
{The aim of this article is to clarify our understanding of the behavior of IGWs in a 3D radiative zone and to provide a clear overview of their
  properties.
}
% Method
{ We use a method of frequency filtering that reveals the path of {individual} gravity waves of different frequencies in
  the radiative zone.
}
% Results
{We are able to identify the region of propagation of different waves in 2D and 3D, to compare them to the linear raytracing theory and to distinguish
  between propagative and standing waves (g modes). We also show that the energy carried by waves is distributed in different planes in the sphere,
  depending on their azimuthal wave number.
}
% Conclusions
{We are able to isolate individual IGWs from a complex spectrum and to study their propagation in space and time. In particular, we highlight
    in this paper the necessity of studying the propagation of waves in 3D spherical geometry, since the distribution of their energy is not
    equipartitioned in the sphere.
}

%=====KEY-WORDS=====%=========%=========%=========%=========%=========%=
\keywords{{Hydrodynamics, Stars: interiors, Sun: interior, Turbulence, Waves}}

%=HEADER==%=========%=========%=========%=========%=========%=
\titlerunning{Characterizing gravity waves in 3D nonlinear stellar simulations}
\authorrunning{L. Alvan, A. Strugarek, A.S. Brun, S. Mathis, R.A. Garcia}

%===MAKETITLE====%=========%=========%=========%=========%=
\maketitle

\section{Introduction}

Internal gravity waves (IGWs) propagate in stably stratified fluids with gravity as a restoring force. 
They can be observed in laboratory experiments 
\citep[e.g.,][]{1967JFM....28....1M,2007ExFl...42..123G,Perrard:2013bc}, planetary atmospheres, and oceans
\citep[e.g.,][]{2002AnRFM..34..559S,ROG:ROG1574}, and in our case, {they are believed to exist} in the radiative regions of stars. In
solar-like stars, these waves are stochastically excited by turbulent convective motions in the outer layers, which leads to the formation of a rich
spectrum. During their propagation, IGWs are known to transport angular momentum by radiative damping
\citep[e.g.,][]{1993AA...279..431S,ZahnTalonMatias1997}, corotation resonances \citep[e.g.,][]{Booker:1967wd,2013AA...553A..86A}, or nonlinear wave
breaking \citep[e.g.,][]{BarkerOgilvie2010}.
They also affect the mixing of chemical elements in stars' interiors \citep[e.g.,][]{Press1981,1991ApJ...377..268G,Montalban:1994wx,CharbonnelTalon2005}. As a
consequence, they are expected to influence the evolution of stars and particularly the evolution of their internal rotation profiles
\citep{Talon:2005iu,Charbonnel:2013df,2013AA...558A..11M}. In the solar case, IGWs are serious candidates for explaining the solid-body
rotation of its radiative interior down to 0.2$R_\odot$ \citep{1999ApJ...520..859K,CharbonnelTalon2005}. 
\footnote{{Another hypothesis for explaining the solid body rotation of the solar radiative zone relies on the existence of a buried large-scale
    fossil magnetic field, whose existence and effect are still the subject of debates in
    the community \citep{GoughMcIntyre1998,2011A&A...532A..34S,Acevedo13}.}} They are also invoked to explain the
  transport of angular momentum in sub- and redgiant stars \citep{2008AA...482..597T,2014arXiv1409.6835F} and in intermediate 
  mass and massive stars \citep{Pantillon:2007cc,2014MNRAS.443.1515L}.\\\newline
When progressive waves interfere constructively, they form standing modes called gravity (g) modes,
whose measurement can provide information about the structure of the stars' deep interiors. {These individual g modes remain barely detectable at the
  surface of the Sun.} 
%Their detection in the Sun is one of the main questions in helioseismology}. 
Unlike acoustic (p) modes, mainly located near the surface, g modes are trapped in the inner radiative region, so we can accumulate
a lot of information about the structure of this zone. Nevertheless, they are evanescent in convective regions and thus reach the
surface with small amplitudes \citep[][and references therein]{2010AARv..18..197A}. {As of today, the only reported detection for the Sun has been done in}
\citet{Garcia:2007iq, GarciaGimenez}. They observed the asymptotic signature of g modes at the
surface of the Sun, but we are not yet able to detect individual peaks \citep{GarciaChimera}. {To improve our ability to detect them, we thus
  need to characterize their properties, time variabilities, and ability to tunnel through the solar convection zone better.}\\\newline
Realistic numerical simulations help for understanding the properties of IGWs in stars. Their excitation by the pummeling of convective plumes at the
top of the radiative zone has already been studied in 2D
\citep{1984AA...140....1M,1986ApJ...311..563H,1994ApJ...421..245H,2003AcA....53..321K,2005AA...438..365D,2005ApJ...620..432R,Rogers:2006ks} and in 3D
\citep{2000ApJ...529..402S,2002ApJ...570..825B,BBT2004,Kiraga:2005wg,Brun:2011bl}. Recently, \citet{Alvan:2014gx} have shown the substantial progress made in
high-performance computing, allowing a 3D solar-like star to be modeled in full spherical geometry from $r=0$ to $r=0.97R_\odot$. They took the nonlinear coupling between the convective envelope, the radiative interior, and the dissipation along the wave propagation into
account. Thanks
to the use of a realistic density stratification in the radiative region,  
the properties and frequencies of the waves excited by the convection are those expected from theory. We seek here to characterize the
  spectrum excited better by doing a detailed analysis of its substructures. More precisely, we want to understand why we can see modes in the
    spectral space and not in the direct observation of the simulation (real space). Our guiding question concerns the distinction between
    progressive and standing waves and the transition between these two families. Finally, we see how the waves' energy is distributed in the
    spherical star and what can be deduced from their excitation.\\\newline
In the present work, we use a method of frequency filtering that reveals the signature of IGWs in real space and that clarifies their behavior at different
frequencies. In section 2, we show and discuss the complex and rich spectrum of IGWs excited by penetrative convection in our 3D model. In section 3,
we describe our method of frequency filtering before presenting the associated results. We thus show the relation between our full 3D nonlinear
simulations and the asymptotic linear raytracing theory, often used to illustrate the 
propagation of internal waves in stars. We also show that waves of low frequency are attenuated while higher-frequency waves propagate up to the
center and form modes. We finish by showing a property of internal waves that single-handedly warrants their study in 3D: at low
rotation rate, the energy carried by IGWs is concentrated in planes of the sphere, whose inclination depends on the wave numbers. This property is
broken at high rotation rates, leading to very complex 3D waves' paths of propagation.

\section{3D nonlinear simulations of IGWs}

{We used the anelastic spherical harmonic (ASH) code \citep{CluneAl1999, BMT2004} to solve the full set of 3D anelastic hydrodynamic
equations in a solar-like star rotating at the velocity $\vec\Omega_0=\Omega_0\vec{e}_z$ {($\Omega_0$ = 414 nHz)}. The domain of
computation includes both the radiative and 
the convective regions from $r=0$ up to $r=0.97R_\odot$. We 
provide a description of the ASH code and a presentation of the model used here in Appendix A. More details can be 
  found in \citet{Alvan:2014gx}. In this section, we focus on the study of the gravity wave spectrum excited in the radiative zone. This spectrum is
  rich and complex, and it varies as a function of depth.}

\subsection{Wave pattern in physical space}

\begin{figure}[h]
  \centering
  \includegraphics[width=0.4\textwidth]{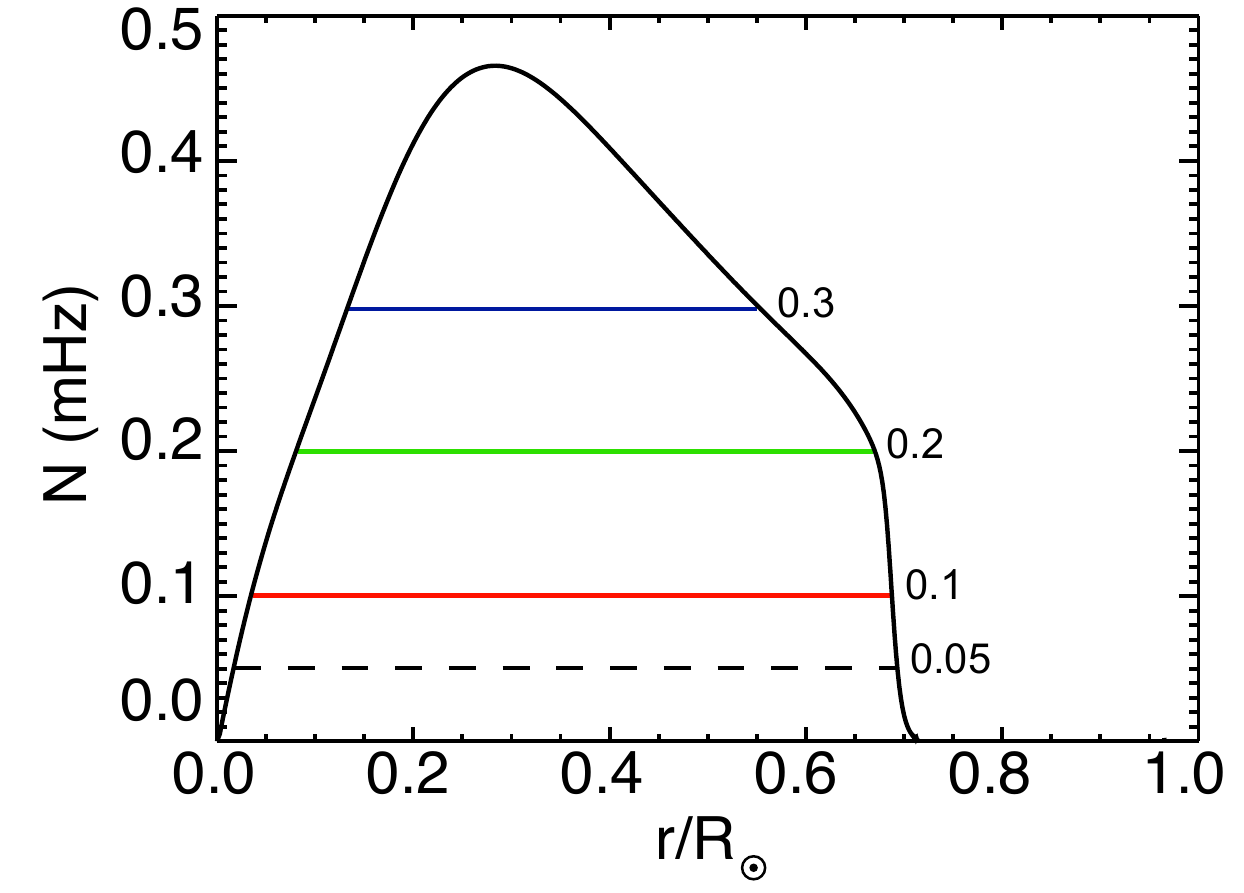}
  \caption{Profile of the Brunt-Väisälä frequency as a function of the normalized radius. Colored horizontal lines highlight the regions 
of propagation of the waves revealed by our filtering technique in Fig. \ref{fig:filtres_reels}.}
  \label{fig:N}
\end{figure}

The use of a seismically calibrated 1D solar model \citep{BrunAl2002} to initialize the 3D simulation ensures a realistic 
stratification in the radiative zone that allows us to study the properties of the IGWs excited by the overshoot process quantitatively.
To quantify this stratification, we define the Brunt-Väisälä frequency by
\begin{equation}
N = \sqrt{\bar{g}\left(\frac{1}{\gamma\bar{P}}\frac{\partial \bar{P}}{\partial r}-\frac{1}{\bar\rho} \frac{\partial \bar\rho}{\partial r}\right)}\hbox{,}
\end{equation}
where $r$ is the radius, $\bar\rho$, $\bar{P}$, and $\bar{g}$ are the reference density, pressure, and acceleration of gravity, and $\gamma$ is the first
adiabatic exponent (see Appendix A). {In the whole paper, frequencies are plotted in Hz and not in rad/s.}
We represent the profile of $N$ in Fig. \ref{fig:N}. {}\\\newline

\begin{figure*}[p]
  \centering
  \includegraphics[width=0.82\textwidth]{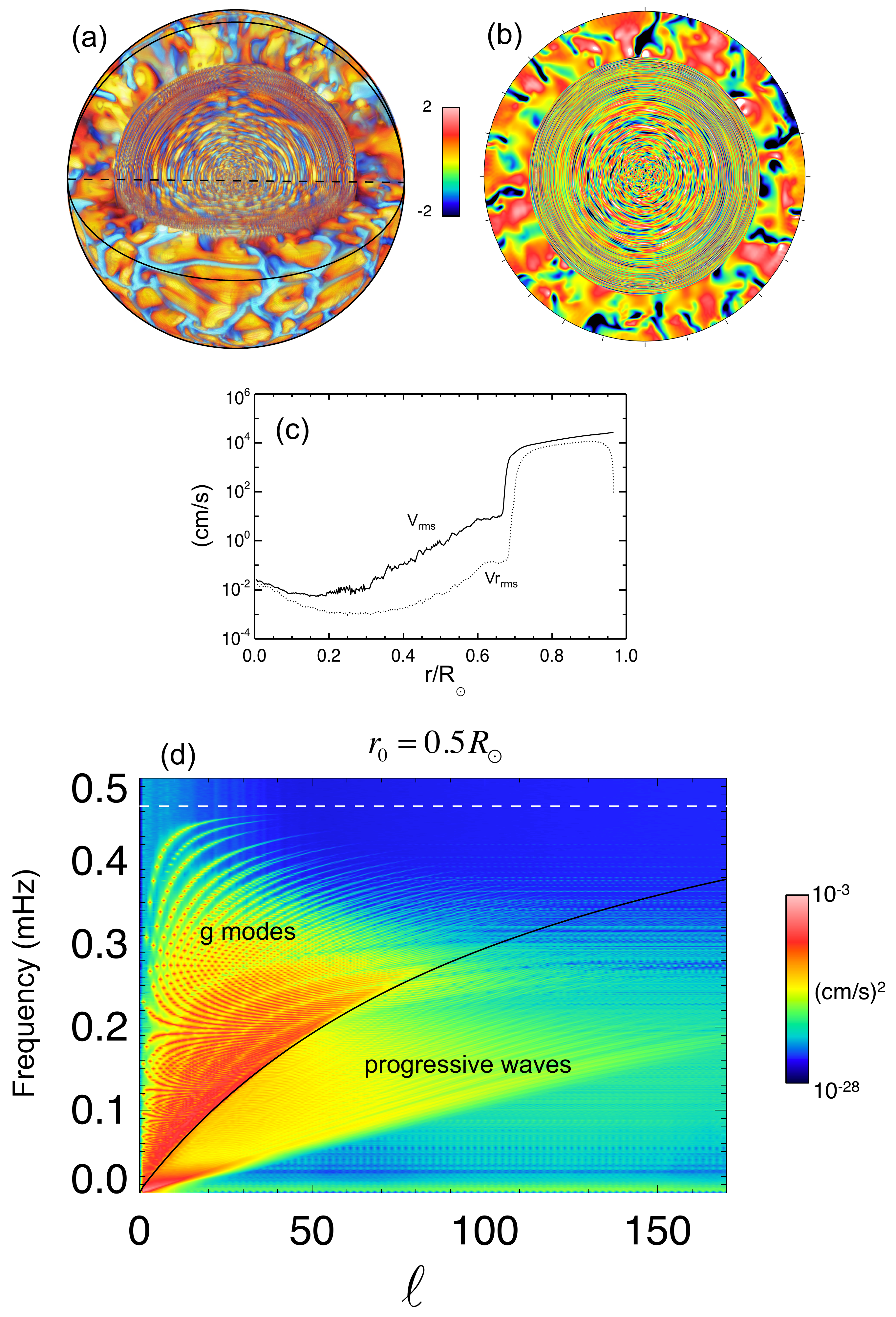}
  \caption{{(a)}: 3D rendering of the simulated star. We represent the radial velocity divided by its rms value at each radius.
{(b)}: Equatorial slice. Like in the 3D view, IGWs pattern are visible in the inner radiative zone as quasi-circular spirals.
{(c)}: rms profiles of the total (solid line) and radial (dotted line) velocities as a function of the
    normalized radius, averaged over longitude, latitude, and time (about 10 convective overturning times).
{(d)}: Energy spectrum of gravity waves computed at $r_0=0.5$ $R_\odot$ as a function of degree $\ell$ and frequency $\omega$. Ridges are
    formed by g modes of same radial order $n$, and we see that they tend to the maximum Brunt-Väisälä frequency at high order $\ell$ (dotted white
    line). The black solid line denotes the separation between g modes (above) and progressive waves (below).}
  \label{fig:Bigfig}
\end{figure*}

The linear asymptotic theory \citep[e.g.,][]{1989nos..book.....U,aerts2010asteroseismology,JCD2011lect} predicts that the dispersion relation of IGWs is
\begin{equation}
  \label{eq:disp}
  \omega = \frac{N k_h}{k}\hbox{,}
\end{equation}
where $\omega$ is the frequency of the wave {(also expressed in Hz),} and $k$ the norm of the wavevector $\vec k = k_r\vec{e}_r + \vec{k}_h$ whose radial and horizontal
components are denoted $k_r$ and $k_h$. This equation is satisfied when we neglect the effect of the Coriolis acceleration, which is possible here
since the model rotates at the solar rotation rate, {verifying $2\Omega_0 = 2 \times 414 $ nHz $\ll \omega$} {for the frequencies $\omega$ of
  interest in this work}. According to this relation, IGWs can propagate in the regions where $N$ is real {only}, i.e. in radiative zones,
and are evanescent in convective zones. Their {maximum} frequency is also limited by the maximum value max$(N)=0.466$ mHz available for the current
Sun (cf. Fig. \ref{fig:N}).\\\newline
In Figs. \ref{fig:Bigfig}(a) and \ref{fig:Bigfig}(b), we show the radial velocity in a 3D view and an equatorial slice of the simulated star with
downflows and upflows. We clearly see the convective envelope from about 0.71$R_\odot$ to the surface of the computational domain and the inner
radiative zone. {Since it is visible in Fig. \ref{fig:Bigfig}(c), the velocity amplitude of the convective motions is much larger than the one of the
  waves (factor $10^3$ to $10^5$). Consequently, to visualize both convection pattern and IGWs in the top panels, we divided the radial velocity by its root mean square (rms) value at each 
radius. }\\\newline
In the radiative region, we see
what looks like concentric circles, which are in fact almost circular spirals. They correspond to the 
superposition of several wavefronts of gravity waves evolving at different frequencies. The comparison between these pattern and those predicted by
the raytracing theory (see Sect. \ref{sec:visu-high-freq}) shows that the IGWs visualized here have frequencies of about [0.01-0.05] mHz. This is quite low in
comparison with the expected solar g modes' frequencies \citep{JCD2011lect}. Deeper in the radiative region, the quasi-circular pattern
  changes into a more complex shape. We explain these observations in Sect. \ref{sec:dist-betw-progr}.

\subsection{Spectrum}

To describe the IGWs we can see into the radiative zone more precisely, we represent their
spectrum so as to calculate their frequencies. It is obtained by applying a spherical harmonic transform, followed by a temporal Fourier transform, to
the radial velocity 
$\mathrm{v}_r(r_0,\theta,\varphi,t)$. For the moment, the radius $r_0$ is fixed, and we obtain the quantity $\tilde{\mathrm{v}}{_r}(r_0,\ell,m,\omega)$,
where $\ell$ is the degree, $m$  azimuthal number, and $\omega$ the frequency of the wave. The quantity represented in Fig. \ref{fig:Bigfig}(d) is
obtained by adding all contributions in $m$ quadratically, such as
\begin{equation}
\label{eq:summ}
  E(r_0,\ell,\omega) = \sum_{m} |\tilde{\mathrm{v}}_r(r_0,\ell,m,\omega)|^2 \hbox{.}
\end{equation}

\begin{figure}[]
  \centering
  \includegraphics[width=0.45\textwidth]{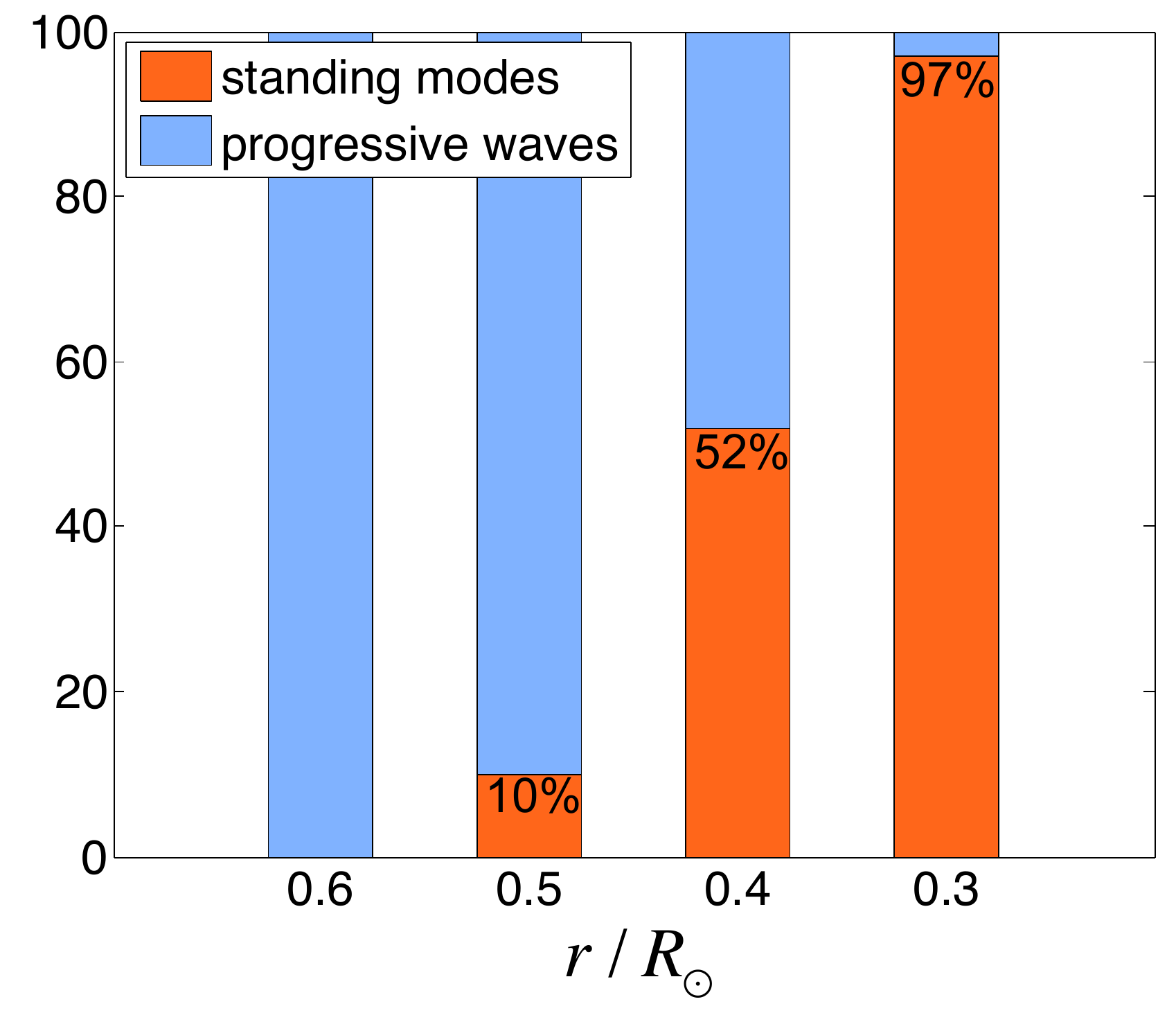}
  \caption{Percentage of the total energy (squared radial velocity) distributed in standing and progressive gravity waves.}
  \label{fig:e_modes}
\end{figure}

% \LA{This limit is lower (black dotted line) for model \textit{turb2$\kappa$}. We also see that the
%   progressive waves zone is largely reduced in comparison with \ref{fig:Bigfig}(d).}

The length of the temporal sequence used here is about 30 days, which provides a frequency resolution of $\omega_{\rm{min}}\approx 0.38 \times 10^{-3}$
  mHz. The temporal sampling rate $\Delta t = 1000$s allows us to reach the maximum (Nyquist) frequency $\omega_{\rm{min}} = 0.5$ mHz and corresponds to
  ten times the time step of the simulation.\\ 

The figure we obtained (Fig. \ref{fig:Bigfig}(d)) is close to the spectrum predicted by the linear theory 
\citep[e.g.,][]{1986AA...165..218P,ChristensenDalsgaard:2003vx}. In the higher part of the spectrum (above the black curve), we see
individual modes forming energy peaks. Their frequencies correspond to g-mode frequencies calculated by the oscillation code
  ADIPLS\footnote{http://users-phys.au.dk/jcd/adipack.n} with  very good accuracy \citep{Alvan:2014gx}.
Modes with the same radial order $n$ - corresponding to the number of zeros of the radial eigenfunctions
- form ridges, particularly visible at high frequency. Moreover, as predicted by the dispersion relation (Eq. \ref{eq:disp}), the frequency upper limit corresponds to
max$(N)=0.466$ mHz. {The lower limit of this ``g modes'' region is the effect of waves attenuation. {It is shown here and can be understood as a cut-off frequency for the formation of g modes, {under which the waves are
      sufficiently damped so that standing modes cannot form.} We show in Appendix B that this  
    curve behaves as $\left[\ell(\ell+1)\right]^{3/8}$.} 

Below this frontier, the spectrum is composed of progressive waves. When we compare spectra
  at different radii $r_0$ \citep{Alvan:2014gx}, we see that the ``progressive-wave region'' decreases with
increasing depth and finally disappears. {We compare the proportion of energy distributed in both regions in Fig. \ref{fig:e_modes}. At
  $r=0.6$R$_\odot$, the whole energy is contained in progressive waves. Then, when we move deeper into the radiative zone, g modes appear and
  progressive waves are damped. Below 0.3R$_\odot$, the number of progressive waves becomes negligible, and the whole energy of the spectrum corresponds to
  standing modes.}
% Moreover, if we change the value of the radiative diffusivity $\kappa$ of our model, the position of the
% black curve changes as well. In Fig. \ref{fig:spectre}, we compare the current model \textit{turb2} with an other model \textit{turb2$\kappa$}, identical in every points except for $\kappa$
%  which is divided by two. The result is that the g mode zone extends beyond the limit (black curve). The new frontier is materialized by the black
%  dotted curve.} We explain these observations by the presence, in our simulation, of radiative effects, that damps the waves.  We will see a further
% explanation in Sect. \ref{sec:dist-betw-progr}.\\

The richness of this spectrum shows that a wide frequency range is excited by convection. As a result, one may wonder why we only see one main pattern
in real space 
(Fig. \ref{fig:Bigfig}(a) and (b)). The raytracing theory predicts that IGWs propagate along different paths depending on their frequencies. Is there a way to
visualize these paths in our simulations? By isolating some waves and visualizing them in the model, we show that we can learn more about
the regions of propagation of IGWs in a full 3D geometry. 

\section{Going further: frequency filtering of radiative zone}

Here, we have filtered our data by selecting only a narrow band of frequencies. In the following sections, we show that this
method reveals important properties of IGWs that were not accessible before.

\subsection{Method}
\label{sec:method}

\begin{figure*}[]
  \centering
  \includegraphics[width=1\textwidth]{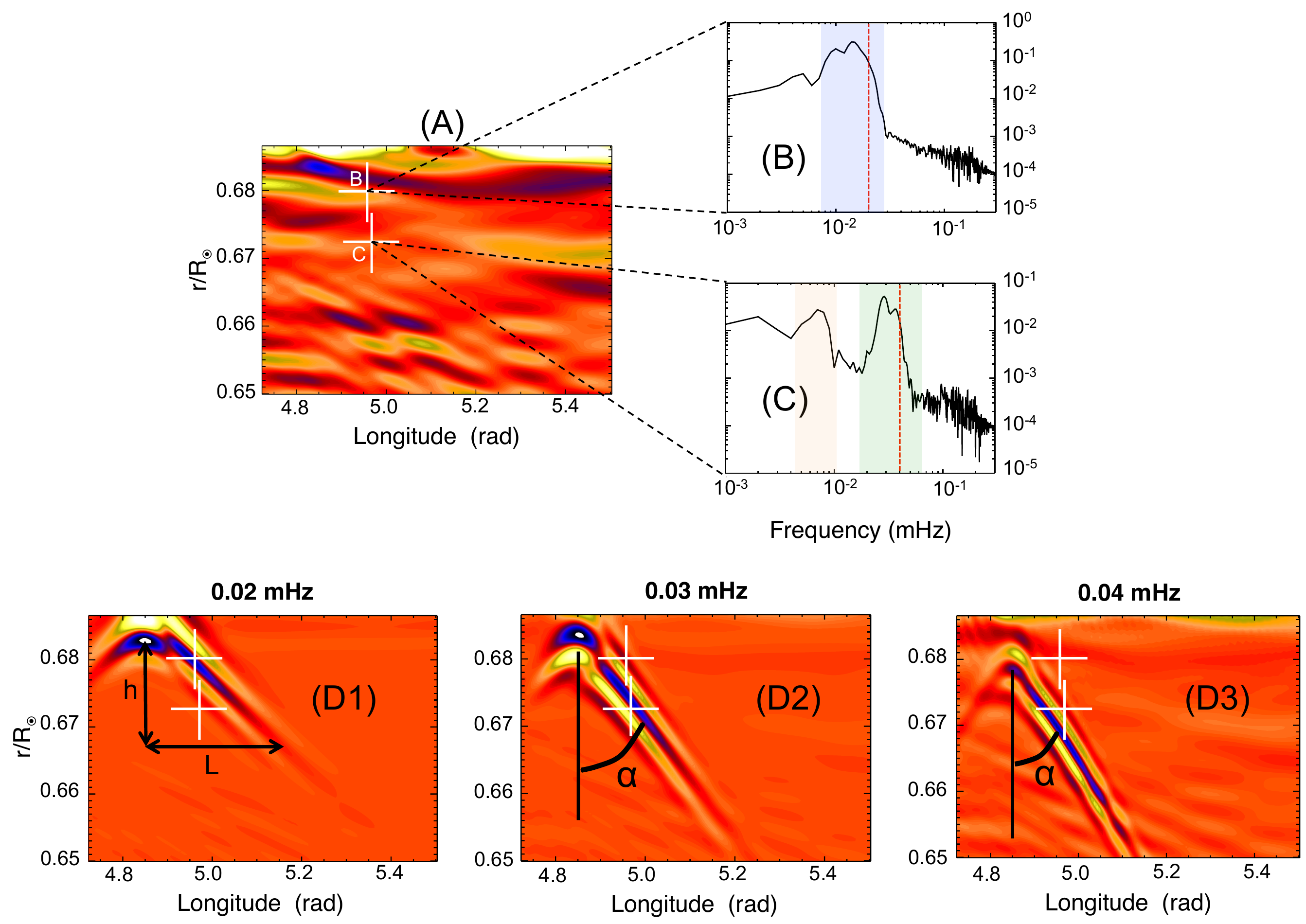}
  \caption{Filtering of the same image at three different frequencies. Downflows are represented in blue, upflows in red tones. {\textbf{(A)}}:
    selected zone of the star. {\textbf{(B)} and \textbf{(C)}} : 
    frequency spectra of points B and C where we see one or two main peaks. The vertical axis represents the normalized radial velocity. The
    vertical red line indicates the position of the Gaussian filter (width = 2$\times$10$^{-4}$mHz). {\textbf{(D1)},
      \textbf{(D2)} and \textbf{(D3)}}: result of the filtering of image \textbf{(A)} at 
    three frequencies. Wave beams are visible with an inclination that varies with the frequency, forming St Andrew's crosses. The angles' values are indicated in Table \ref{tab:chap6_angles}.}
  \label{fig:expli_filtrage}
\end{figure*}

The idea of the frequency filtering can be discussed as follows. We illustrate the method in Fig. \ref{fig:expli_filtrage}.\\\newline
Panel (A) shows a region of the simulated star belonging to the equatorial plane. We represent the radial velocity as a function of the
  radius and the longitude, right below the excitation zone (interface between convective and radiative regions). Several different gravity waves are
  propagating through this region, but they are placed on top of each other so we cannot see their paths clearly. Our aim is to separate the waves of
  different frequencies. To illustrate the manipulation, we have chosen two points (B and C) labeled by white crosses. Image (A) evolves with time
  and so does the signal at points B and C. \\\newline
Panels (B) and (C) represent the Fourier spectrum of these two points. For Point B, we see
  that the main peak is located in the blue shaded region, around 0.01-0.02 mHz. For Point C, the main peak is instead around 0.03 mHz (green shaded region) but we
  also see a secondary peak around 0.005 mHz (orange shaded region). This means that each point oscillates at the frequency of one or two waves passing
  through. \\\newline
Thus, to isolate a wave oscillating at frequency 0.02 mHz (for example), we calculated the Fourier spectrum of each point (i.e. $N_r\times
  N_\theta\times N_\varphi=569 \times 512 \times 1024 \approx 3\times 10^{8}$ points), we applied a passband filter to the spectrum (a Gaussian centered
  at 0.02 mHz), and we came back to the real space using an inverse Fourier transform. The width of the Gaussian filter is $2\times 10^{-4}$ mHz for
  this example. \\\newline
The result is shown in Panels (D1), (D2), and (D3) of Fig. \ref{fig:expli_filtrage} for three different frequencies. For 0.02 mHz, we see that the
isolated wave passes through Point B, and it is also the case for the wave at 0.03 mHz passing through Point C. We note $\alpha$ the angle
  between the direction of the wave beam and the gravity. }This angle decreases from (D1) to (D3), since we increase the central frequency of
the filter.

  \begin{table}[h]
 \centering
\begin{tabular}[h]{|c|c|c|c|}\hline
    \centering
    Filtered frequency & $N_1$ (mHz) & $\alpha_{\rm{th}}$ (deg) & $\alpha_{\rm{m}}$ (deg)\\\hline
    0.02 mHz & 0.15 $\pm$ 0.02 & 82.34 $\pm$ 1.03& 85.4 $\pm$ 2.29  \\
    0.03 mHz & 0.17 $\pm$ 0.02 & 79.84 $\pm$ 1.21 & 84.29 $\pm$ 3.59 \\
    0.04 mHz & 0.18 $\pm$ 0.02 & 77.16 $\pm$ 1.45 & 82.2 $\pm$ 1.79 \\\hline
\end{tabular}
    \caption{Comparison between the St Andrew's cross angles predicted by the dispersion relation and measured in Fig. \ref{fig:expli_filtrage}. Owing to the zooming effect, the $\alpha$ angles actually look smaller in the lower panels of Figure 4.} 
    \label{tab:chap6_angles}
  \end{table}

\begin{figure*}[]
  \centering
  \includegraphics[width=0.75\textwidth]{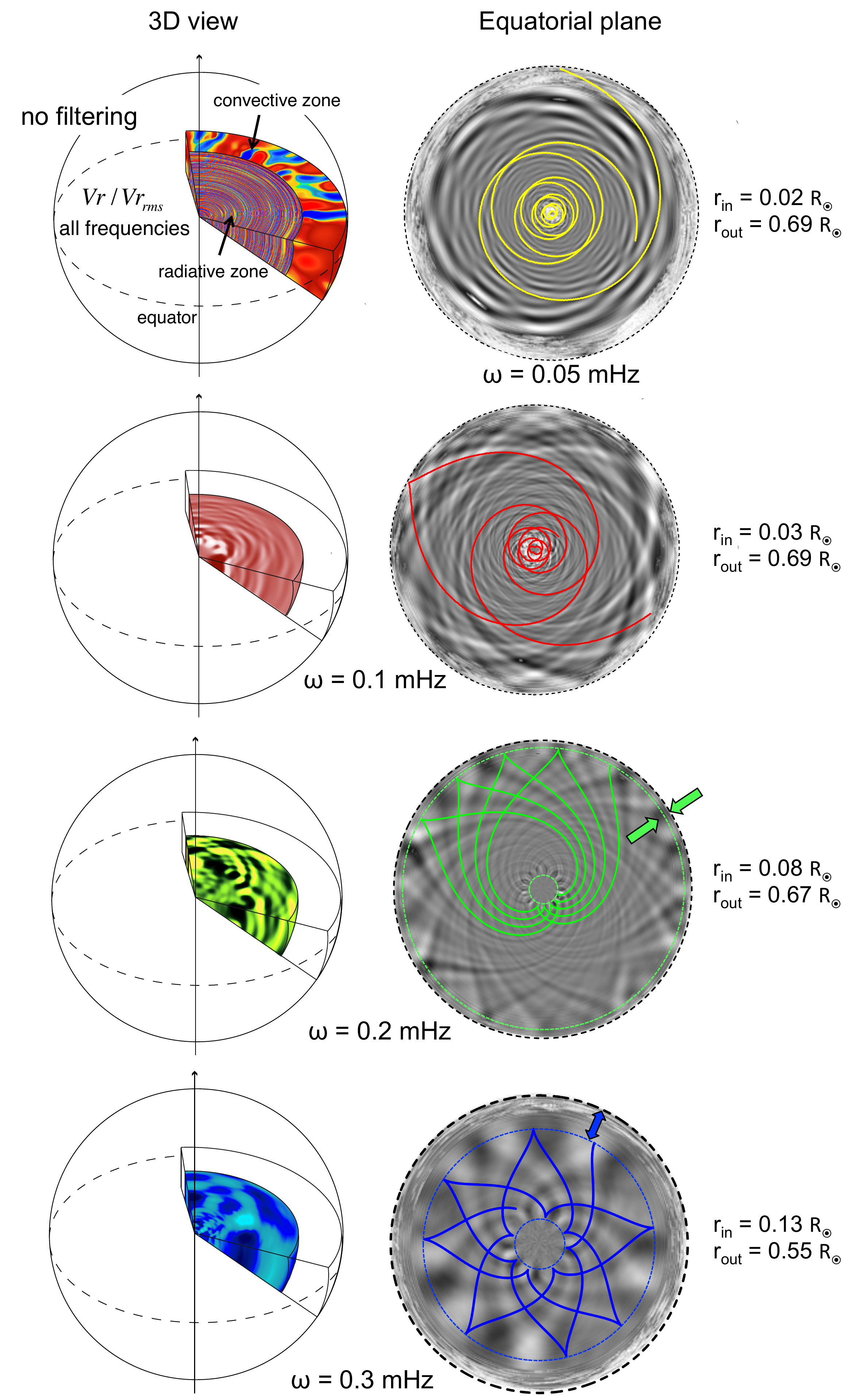}
  \caption{3D and 2D views of portions of the star filtered at different frequencies. On the right, ASH results are represented in gray tones, and we
    have superimposed the path of the 
    corresponding wave calculated by the method of raytracing. Only the radiative zone is represented. For $\omega=0.2$ mHz, the green dotted circle
    represents the outer turning point, and the 
    green arrows highlight the nonpropagation region. For
    $\omega=0.3$ mHz, the two blue dotted circles represent the
    outer and inner turning points, and the blue arrow shows the nonpropagation region. For the four frequencies, the propagation regions correspond to the colored horizontal lines in Fig.~\ref{fig:N}.}
  \label{fig:filtres_reels}
\end{figure*}

This manipulation allows us to highlight one of the main properties of IGWs. From the linear theory of raytracing
\citep{1978cup..book.....L,goughHouches,JCD2011lect}, we know that IGWs propagate
along beams whose paths are ruled by the dispersion relation (Eq. \ref{eq:disp}). The shape formed by the beams at the point of the wave's
excitation is called a St Andrew's cross \citep{LighthillBook,1991JFM...231..439V}. The wave's energy is radiated around
an angle $\alpha_{\rm{th}}$ to the radial direction such that 
\begin{equation}
\label{eq:4}
  \alpha_{\rm{th}} = \arccos\left({\frac{\omega}{N_1}}\right)\hbox{,} 
\end{equation}
where $N_1$ is the value of $N$ in the region considered. This angle is visible in Panels (D1), (D2), and (D3) of
Fig. \ref{fig:expli_filtrage}.  To measure it, we calculate its tangent in each panel using the relation
\begin{equation}
  \label{eq:tana}
  \alpha_{\rm{m}}=\arctan\left(\frac{L}{h}\right)\hbox{,}
\end{equation}
where $L$ is the length considered along the longitudinal direction and $h$ along the radial direction.
For the measurement of $L$, we have translated the longitude from radians to solar radii. The values found are indicated in
Table \ref{tab:chap6_angles} and compared to pseudo-theroretical values calculated from Eq. \eqref{eq:4}, {where the value of $N_1$ must be
  deduced from the figure.}\\\newline
For $\alpha_{\rm{m}}$, the main source of error is the measurement of the distances $h$ and $L$. The error shown here corresponds to the width
  of the beam. For $\alpha_{\rm{th}}$, we also considered an error coming from the measure of $N_1$, which depends on the estimation of the starting
  point of the cross. Another possible source of error could be the quality of the filter, but since it has a very narrow bandwidth, this is negligible
  compared to the other factors. \\

{This comparison shows that we find a bias between both values, where $\alpha_m$ is systematically larger. A future work will be dedicated to exploring more frequencies, in
  order to determine the extent of this difference precisely and to decide on the precision of the linear dispersion relation used here. We have already
  checked that the influence of the coriolis acceleration is too weak at this rotation rate to explain the observed difference, even though it goes in
  the right direction.} 

\subsection{Visualization of the high frequency wave pattern}
\label{sec:visu-high-freq}

\begin{figure}[h]
  \centering
  \includegraphics[width=0.5\textwidth]{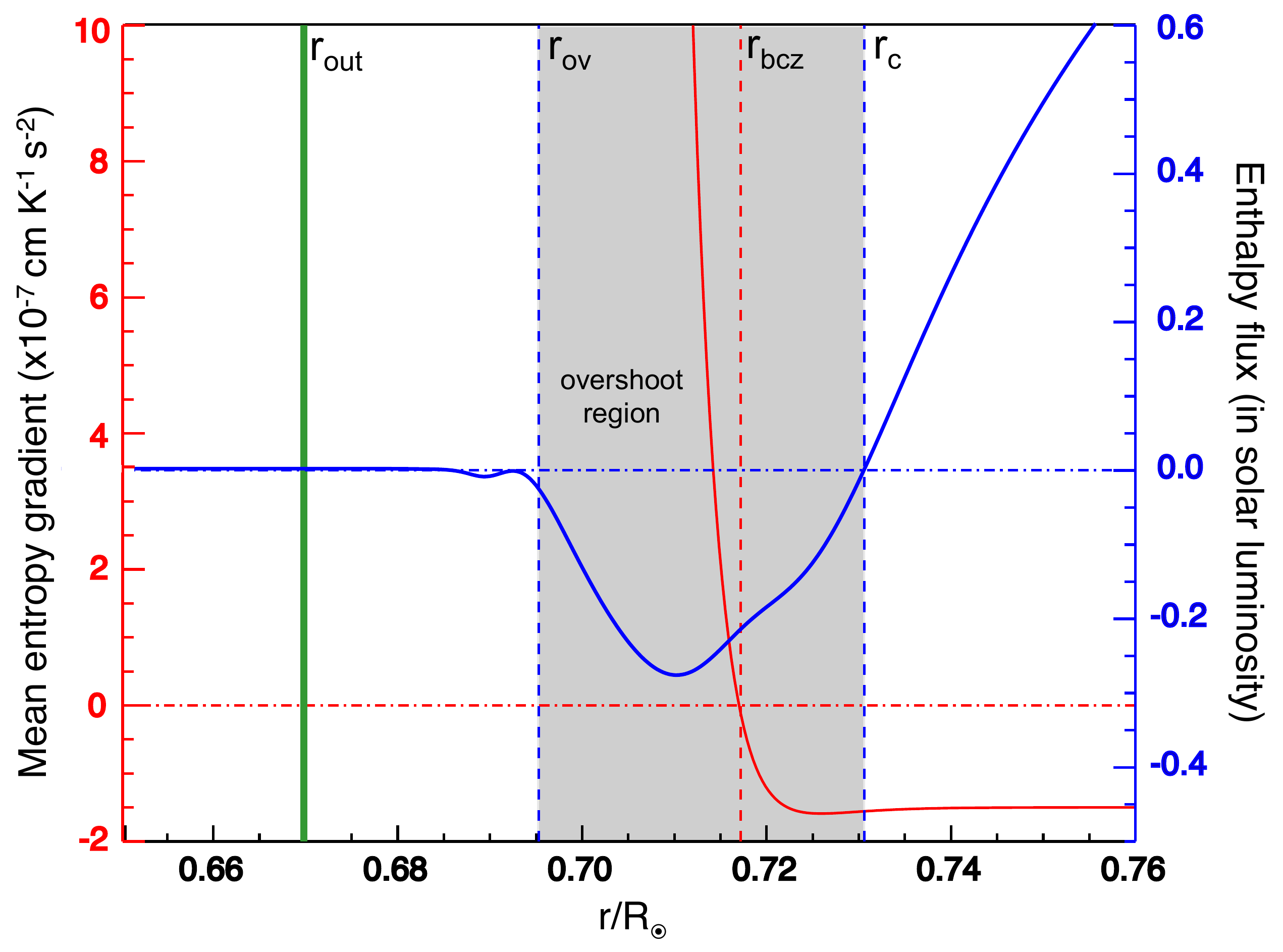}
  \caption{Representation of the overshoot region defined between $r_{\rm{c}}$, the first zero of the enthalpy flux, and the point $r_{\rm{ov}}$ below which the
amplitude of the flux drops to 10\% of its most negative value. The radius $r_{\rm{bcz}}$ corresponds to the point where the mean entropy gradient changes sign.
 The green vertical line represents the positions of $r_{\mathrm{out}}$ for $\omega$ = 0.2 mHz.}
  \label{fig:schema_rayons}
\end{figure}

We now apply this method of filtering to a larger portion of the star to see if we can visualize other patterns than the quasi-circular spiral visible in
Fig. \ref{fig:Bigfig} (a) and (b). Our results are shown in Fig. \ref{fig:filtres_reels}, in 3D and 2D views. To understand the results, we
computed the paths in 2D of some IGWs using the method of raytracing. It consists in calculating the value of the wave's phase along a path
$\vec{x}(t)$ by resolving the Hamiltonian system  
\begin{equation}
  \label{eq:hamil_cart}
  \left\{
      \begin{array}{l}
\displaystyle\frac{\mathrm{d}x_i}{\mathrm{d}t}=\displaystyle\frac{\partial W}{\partial k_i}\hbox{,}\\
\\
\displaystyle\frac{\mathrm{d}k_i}{\mathrm{d}t}=-\displaystyle\frac{\partial W}{\partial x_i}\hbox{,}\\
      \end{array}
  \right.
\end{equation}

where $W(\vec{x},\vec{k},t)=\omega,$ and $(x_i,k_i)$ the cartesian coordinates of the position vector $\vec{x}$ and the wavevector $\vec{k}$
\citep[e.g.,][]{Lignieres:2011ev}. In our case, we employed polar 
coordinates, transforming the system \eqref{eq:hamil_cart} into
\begin{equation}
  \label{eq4:hamil_pol}
  \left\{
      \begin{array}{ll}
\displaystyle\frac{\mathrm{d}r}{\mathrm{d}t}  = \displaystyle\frac{\partial W}{\partial k_r}  = \mathrm{v}_{\mathrm{g}r} \hbox{,}\\
\\
     \displaystyle\frac{\mathrm{d}\theta}{\mathrm{d}t} = \displaystyle\frac{1}{r}\frac{\partial W}{\partial k_\theta}  =
     \displaystyle\frac{\mathrm{v}_{\mathrm{g}\theta}}{r} \hbox{,}\\
\\
     \displaystyle\frac{\mathrm{d}k_r}{\mathrm{d}t}  = -\displaystyle\frac{\partial W}{\partial
       r} {{+\displaystyle\frac{\mathrm{v}_{\mathrm{g}\theta}}{r}k_\theta}} \hbox{,}\\
\\
     \displaystyle\frac{\mathrm{d}k_\theta}{\mathrm{d}t}  =
    -\displaystyle\frac{1}{r}\displaystyle\frac{\partial W}{\partial\theta}  {{-
     \displaystyle\frac{\mathrm{v}_{\mathrm{g}r}}{r}k_\theta}}  \hbox{,}
      \end{array}
  \right.
\end{equation}
where $\vec{\mathrm{v}}_{\mathrm{g}} = \mathrm{v}_{\mathrm{g}r} \, \vec{e}_r + \mathrm{v}_{\mathrm{g}\theta} \, \vec{e}_\theta$ is the group velocity
of the wave at which the ray propagates. In this work, we use the raytracing theory in 2D, as employed by many other authors
\citep[e.g.,][]{Lignieres:2011ev,JCD2011lect,Kosovichev:2011gx}, as a tool for understanding our simulations. {When computing the ray paths
  shown in Fig. \ref{fig:filtres_reels}, we used the Brunt-Väisälä
  frequency profile presented in Fig. \ref{fig:N}.} Further developments of this method
and its generalization in 3D will be the object of a future work (Mathis et al. 2015, in prep).\\\newline
The results shown in Fig. \ref{fig:filtres_reels} illustrate the diversity of the paths followed by the different waves. The lower the frequency,
the more the ray paths look like spirals. Another way to understand it is to say that we retrieve the cross shape at the point of departure of the waves, whose angle $\alpha$ with the radial
direction decreases with increasing frequency. In particular, the yellow case ($\omega=0.05$ mHz) is the closest to the general pattern
(Fig. \ref{fig:Bigfig} (a) and (b)). We thus confirm that what we see without filtering {in the external region of the radiative zone} is a
sample of low-frequency waves. The other paths were not 
visible in Fig. \ref{fig:Bigfig} (a) and (b) because the low-frequency part of the spectrum is dominant in amplitude (red
tones in Fig. \ref{fig:Bigfig} (d)). By filtering out this part, we can visualize the region of propagation of the other
waves, and especially of the resonant g modes. \\\newline
The dispersion relation (Eq. \ref{eq:disp}) imposes that the wave frequencies $\omega$ are less than the Brunt-Väisälä frequency $N$. 
{Consequently, we can define the limit of the propagation region by two turning points $r_{\rm{in}}$ and $r_{\rm{out}}$ such that
  $N(r_{\rm{in}})=N(r_{\rm{out}})=\omega$.} For instance, the wave oscillating at 0.3 mHz (bottom panel of 
Fig. \ref{fig:filtres_reels}) is confined more deeply in the radiative zone than the one at 0.05 mHz (top panel). This is also visible in
  Fig. \ref{fig:N}, where we indicate the regions of propagation of the waves shown in this section by colored lines. {We note that the
  outer turning point is different from the penetration depth $r_{\rm{ov}}$, where convective plumes deposit the energy 
  transferred to waves. We clarify the different radii defined here in Fig. \ref{fig:schema_rayons}.} \\\newline

\begin{figure*}[]
  \centering
  \includegraphics[width=1\textwidth]{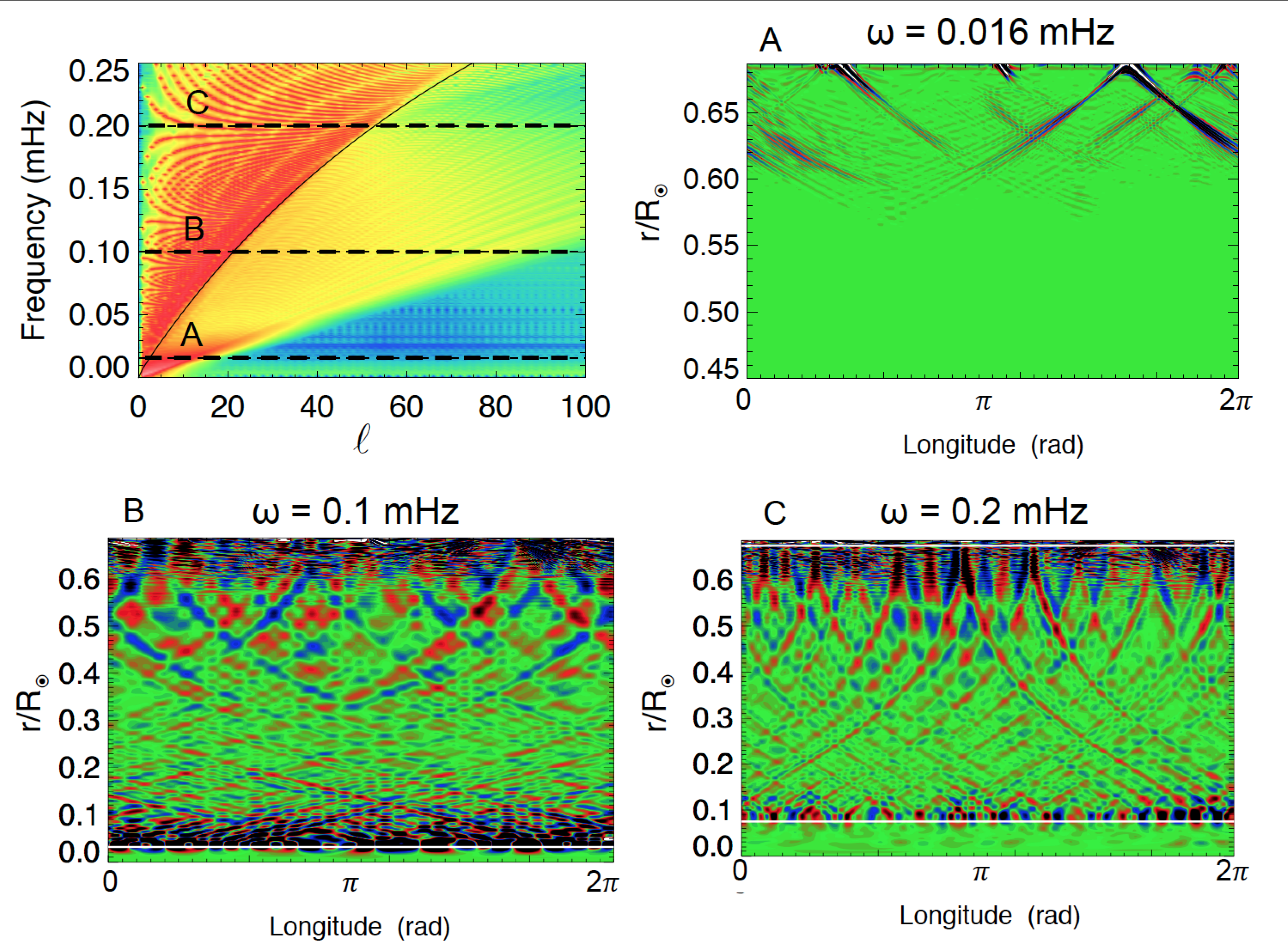}
  \caption{Distinction between progressive and standing waves. The top left panel is a zoom of Fig. \ref{fig:Bigfig}(d) (same colorbar). Three
    horizontal dotted lines represent the frequencies chosen in Panels A, B, and C. In these panels, we show the radial velocity in the same region filtered at three
    different frequencies. Low-frequency waves are rapidly damped and cannot form g 
    modes (Panel A). In contrast, for $\omega=0.1$ mHz and $\omega=0.2$ mHz (Panels B and C), we see modes oscillating 
    between the two turning points (white lines).}
  \label{fig:verte}
\end{figure*}

It is interesting to see that, although
the excitation region is located at the base of the convection zone (penetration region), the perturbation is able to propagate in a ``non-propagation region''
(blue arrow and limits materialized by green and blue dotted circles). This phenomenon can explain the fact that high-frequency modes ([0.3-0.4] mHz) are
less powerful in the spectrum (Fig. \ref{fig:Bigfig} (d)). In fact, a part of the initial energy transfered from the convective region to the waves is lost in the
evanescent region, {between the overshoot region and $r_{\rm{out}}$}.\\\newline %, proportionnaly to} \\$e^{-\displaystyle\int_{r_{\rm{out}}}^{r_c}{|k_r|\mathrm{d}r}}$,
%with $k_r=\left(\displaystyle\frac{N^2}{\omega^2-1}\displaystyle\frac{\sqrt{\ell(\ell+1)}}{r}\right)$.\\\newline
A moving version of Fig. \ref{fig:filtres_reels} shows that the paths visible here do not propagate. They oscillate at the chosen frequency (wavefronts propagate
toward the surface at the phase velocity), but the envelope remains stable. Thus, what we see here are stationary modes instead of progressive
waves. In the following section, we show that the method of frequency filtering allows us to distinguish between these two families.

\subsection{Distinguishing between progressive and standing waves}
\label{sec:dist-betw-progr}

Along their propagation, IGWs are affected by a damping process that is proportional to the radiative diffusivity of the
fluid. {In the linear and asymptotic ($\omega \ll N$) approximations \citep{ZahnTalonMatias1997}}, the amplitude of a gravity wave propagating in a nonadiabatic
medium is damped by a factor $e^{-\tau/2}$ where 
\begin{equation}
  \label{eq:tau}
  \tau\left(r,\ell,\omega\right) =
  \left[\ell(\ell+1)\right]^{\frac{3}{2}}\int_{r}^{{\hat r}_{\rm
      out}}\kappa\frac{N^3}{\omega^4}\frac{\mathrm dr'}{r'^3} \hbox{.}
\end{equation}
We see that this depends on the radiative diffusivity coefficient $\kappa$, but also on both the frequency $\omega$ and the degree
$\ell$. {We introduce ${\hat r}_{\rm out}$, the maximum radius, close to the external turning point $r_{\rm out}$ (with ${\hat r}_{\rm out}<{r}_{\rm out}$), until when the JWKB approximation used to derive this expression can be assumed (see Fig. \ref{fig:schema_rayons} and the detailed discussion in appendix B and in \citet{2013AA...553A..86A}).}\\
%with $r_{ov}$ the radius where the convective plumes deposit their energy to excite waves 
%This expression was obtained by \citet{ZahnTalonMatias1997} in the 
%linear and asymptotic ($\omega \ll N$) approximations. 
%We see that it depends on the radiative diffusivity coefficient $\kappa$, but also on both the frequency $\omega$ and the degree
%$\ell$.\\\newline 
% {We note that, taking into account nonlinear effects, \citet{Alvan:2014gx} showed that this expression surestimates the wave
%   attenuation. Here, we will not discuss this point further, but we will see that the filtering allows to visualize the effect of this damping in the
%   real space.}

We know that g modes are formed by positive interferences between two progressive IGWs. This implies that the amplitude of these IGWs is high enough to
reach the inner turning point $r_{\rm{in}}$ and then to go back toward the surface {(see Eq. \ref{eq:criterion})}. \\\newline 
In Fig. \ref{fig:verte} we show the result of different filtering of the same region of the star. This is a slice of the equatorial plane
($\theta=\pi/2$) with the normalized radius as vertical axis and the 
longitude $\varphi$ as horizontal axis. {In the top righthand panel, we zoom in to the lower region of the spectrum presented in
  Fig. \ref{fig:Bigfig}. We see that, at $\omega=0.016$ mHz, the energy is fairly concentrated below the black line. The result is that the waves
  propagate over a finite distance before being completely damped. We draw the reader's attention to the vertical axis that stops at 0.45 $R_\odot$
  since no wave is visible beyond. In contrast, for $\omega=0.1$ mHz and $\omega=0.2$ mHz, the main energy of the spectrum is above the black line
(red tones), and in the real space, we clearly see that waves propagate between both turning points (white lines). They are excited at the top of the
radiative zone, they go from $r_{\rm{out}}$ to $r_{\rm{in}}$ where they bounce, and they come back to $r_{\rm{out}}$.} \\\newline
{We also note that the path of propagation changes between panels. In particular, the inclination of the rays steepens with higher frequency 
following the tendency discussed in Sect. \ref{sec:method}. Moreover, we see that the number of rays increases with $\omega$. We can define an
effective wavenumber $\ell_{\rm{eff}}$ by}
\begin{equation}
  \label{eq:5}
  \ell_{\rm{eff}} = \frac{2\pi r_{\rm{out}}}{\lambda},
\end{equation}
{where $r_{\rm{out}}$ is the outer turning point, and $\lambda$ a wavelength that can be measured in the figures. We find $\ell_{\rm{eff}} \approx 11$
for $\omega=0.1$mHz and $\ell_{\rm{eff}}\approx 14$ for $\omega=0.2$mHz.}\\\newline
The conclusion of this study is
to confirm the intuition that the spectrum {realized in our 3D simulations} is made up of both standing modes and progressive waves,
{as one expects}. Since the 
radiative damping is more efficient for high values of $\ell$, the corresponding waves do not have enough energy to bounce at
{their inner turning point} and to form modes. They thus stay simple progressive IGWs. {Although not discussed here in detail, it is clear
  that viscosity will also damp the waves (in addition to radiative diffusion), since we have a Prandtl number close to one in the radiative zone \citep{2005JGRD..11015103V}.} 

\subsection{Energy concentrated in planes}
\label{sec:energy-conc-plan}

\begin{figure}[]
  \centering
  \includegraphics[width=0.5\textwidth]{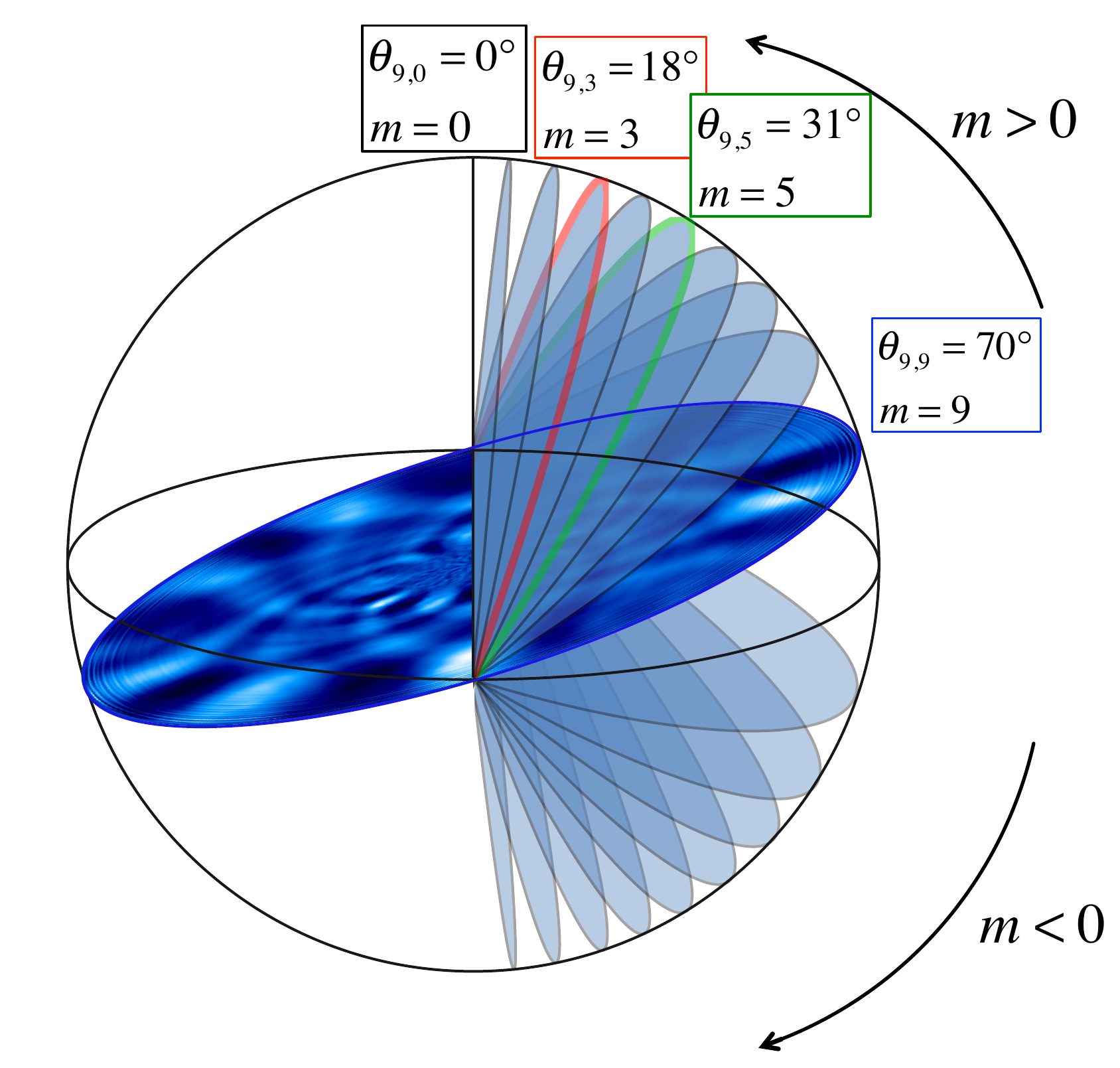}
  \caption{Schematic representation of the propagation planes of azimuthal components of mode $\ell=9$.}
  \label{fig:libellule}
\end{figure}

We now focus on a propagation property of IGWs in 3D. The results of this part highlight the importance of studying gravity waves in fully spherical
geometry. The asymptotic theory of stellar oscillations \citep[e.g.,][]{goughHouches,ChristensenDalsgaard:2003vx,aerts2010asteroseismology} predicts that in the
case where we can neglect the rotation of the star with respect to the waves frequencies ($\omega\ll 2\Omega_0$), the horizontal components of the
wavevector verify
\begin{equation}
  \label{eq:hamil_cart}
  \left\{
      \begin{array}{l}
r\sin\theta\, k_\varphi = m\hbox{,}
\\
k_h^2=k_\theta^2+k_\varphi^2=\displaystyle\frac{\ell(\ell+1)}{r^2}\hbox{.}
      \end{array}
  \right.
\end{equation}
Using this property, \citet{goughHouches} has shown theoretically that the ray paths of waves defined by ($\ell,m$) are confined in planes forming an angle
\begin{equation}
  \label{eq4:thetalm}
  \theta_{\ell,m} = \sin^{-1}\left(\frac{m}{\ell+1/2}\right) \approx \sin^{-1}\left(\frac{m}{\sqrt{\ell(\ell+1)}}\right)\hbox{}
\end{equation}
 with the polar plane.\footnote{The exact expression found by \citet{goughHouches} is the one using $\ell+1/2$, because he employed a general formalism
 that is different from the usual projection on spherical harmonics followed by the asymptotic approximation. Nevertheless, with respect to the precision of
 our results, the approximation $\sqrt{\ell(\ell+1)}\approx 
 \ell+1/2$ is largely acceptable for the degree of the mode chosen here, $\ell=9$.}  
These planes do not depend on the frequency of the waves, {and they are independent of the dispersion relation. For instance, this means that
  this property is also true for acoustic waves.} We show an example in Fig. \ref{fig:libellule}, where we have represented the planes
where modes ($\ell=9,m={0,3,5,9}$) are expected to be concentrated. We see that the highest values of $m$ correspond to planes close to the
equator. By definition, the $m=0$ plane always contains the poles (axisymmetric case).

\begin{figure}[h]
  \centering
  \includegraphics[width=0.5\textwidth]{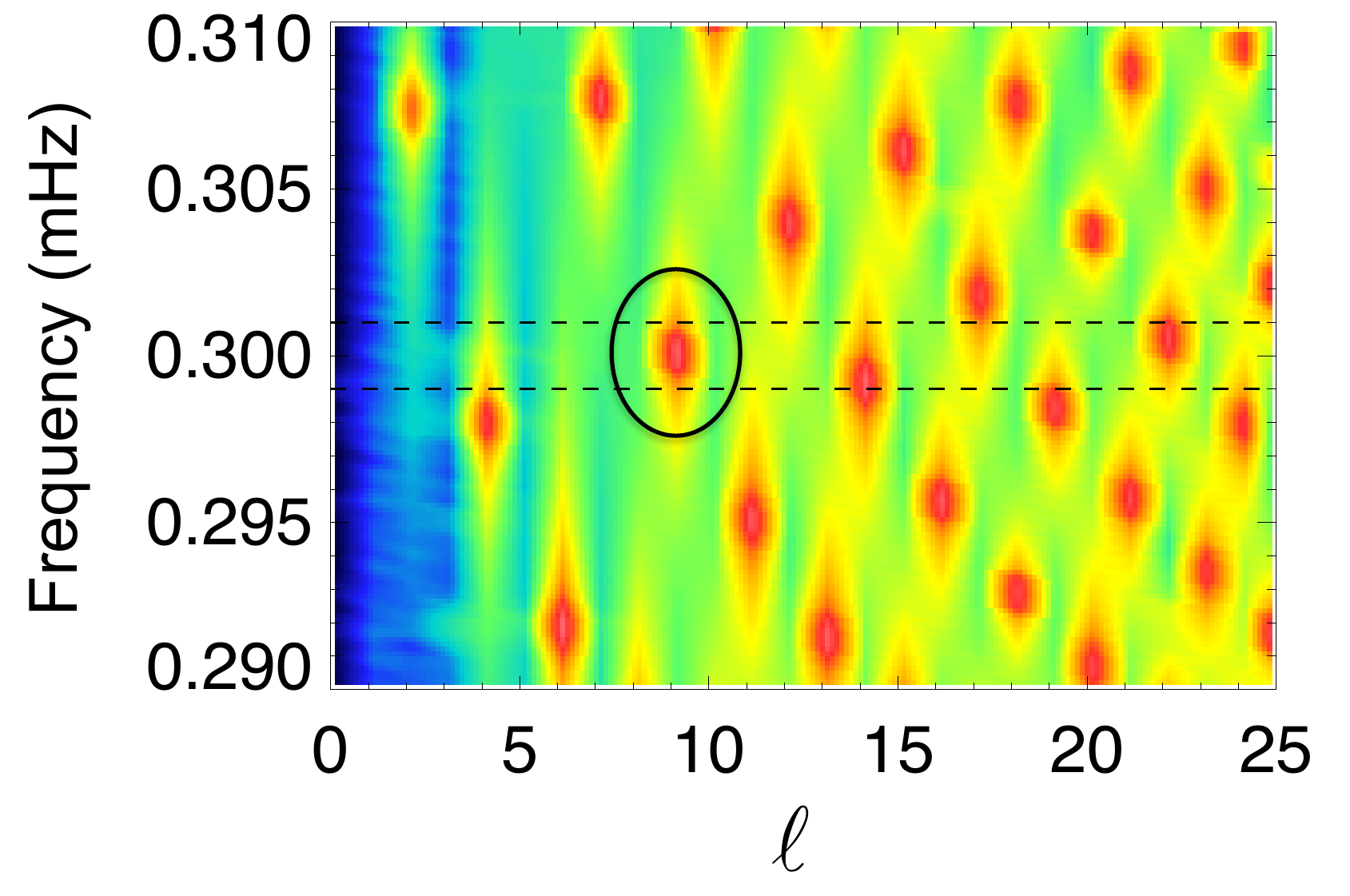}
  \caption{Zoom of the spectrum represented in Fig. \ref{fig:Bigfig}(d). The dotted lines represent the bandwidth of the filter applied
    (Gaussian function centered at 0.3 mHz with full width at half maximum $\sigma=2\times10^{-3}$ mHz) that allows us to select the mode $\ell=9$ in particular.}
  \label{fig:zoom_plans}
\end{figure}

We are going to test this theoretical result with our 3D nonlinear simulations. To illustrate it, we use the result of a filtering at frequency 0.3 mHz. The
corresponding region in the spectrum is represented in Fig. \ref{fig:zoom_plans}. We show a zoom of Fig. \ref{fig:Bigfig}(d) and
indicate the bandwidth of the filter. The mode located in the middle of this bandwith is $\ell=9$. We here recall that the spectrum shown includes all values of
$m$ (see Eq. \ref{eq:summ}). As a consequence, the red peaks of energy visible in this figure are formed by the superposition of all $m$ components,
whose frequency $\omega_{n,\ell,m}$ is slightly shifted by the effect of the rotation. {This rotational splitting is given by}
\begin{equation}
  \label{eq:6}
  \omega_{n,\ell,m}=\omega_{n,\ell,0} + \frac{m}{\ell(\ell+1)}\Omega_0\hbox{,}
\end{equation}
where $\omega_{n,\ell,0}$ is the central peak, corresponding to $m=0$ (see Sect. 4.3.2 of \citet{Alvan:2014gx}). \\\newline
Knowing the rotation rate of our model
(1$\Omega_\odot$) and the frequency considered here, the distance $\omega_{n,\ell,m}-\omega_{n,\ell,0}$ between two peaks of different $m$ and same $\ell$ is much less than the one between
two modes of consecutive $n$ values (see \citet{Alvan:2014gx}). That is why we do not see any overlap in the peak and why the filter
selects the whole azimuthal components of the mode.

\begin{figure*}[]
  \centering
  \includegraphics[width=1\textwidth]{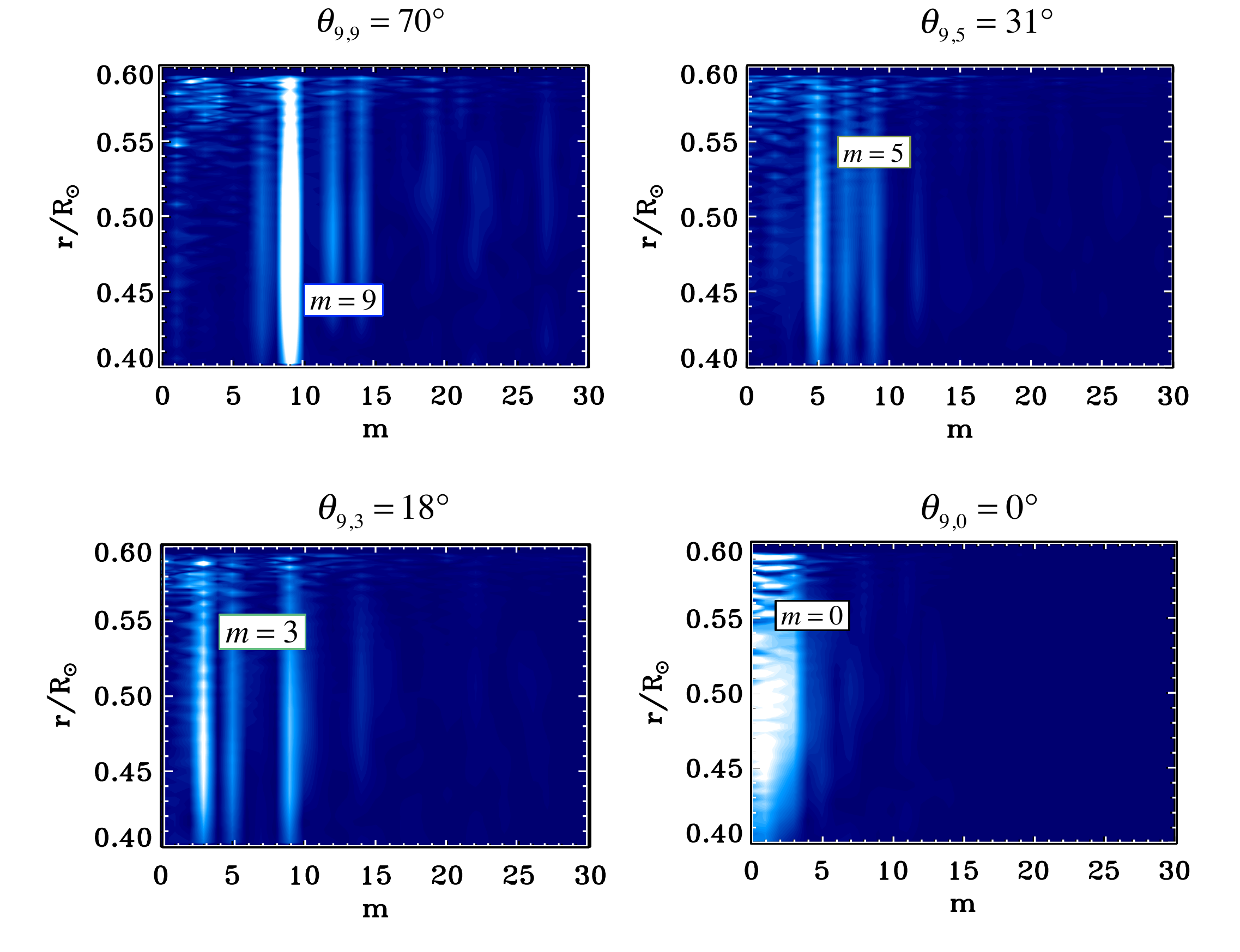}
  \caption{Results of Fourier transforms applied to four selected planes of the radiative zone, after a frequency filtering at $\omega=0.3$ mHz. The
    main value of $m$ in each plane corresponds to the one expected by the asymptotic theory. The horizontal axis is voluntarily extended up to $m=30$ to
    show that we only retrieve the expected values of $m$.}
  \label{fig:fourier}
\end{figure*}

After having filtered the whole radiative zone, we need to access each plane of the sphere and to measure the corresponding dominant azimuthal
number $m$. For example, to reach the plane $\theta_{9,5}=31°$ (represented in green in Fig. \ref{fig:libellule}), we apply a rotation matrix of angle
$90°-\theta_{9,5}$ to the sphere, so that the desired plane is now located at the equator.  We then apply a Fourier transform with respect to the
longitude $\varphi$ to obtain the spectrum in $m$. The result for four planes is given in Fig. \ref{fig:fourier}. The white zones correspond to the
dominant values of $m$.\\

\begin{figure*}
  \centering
  \includegraphics[width=1\textwidth]{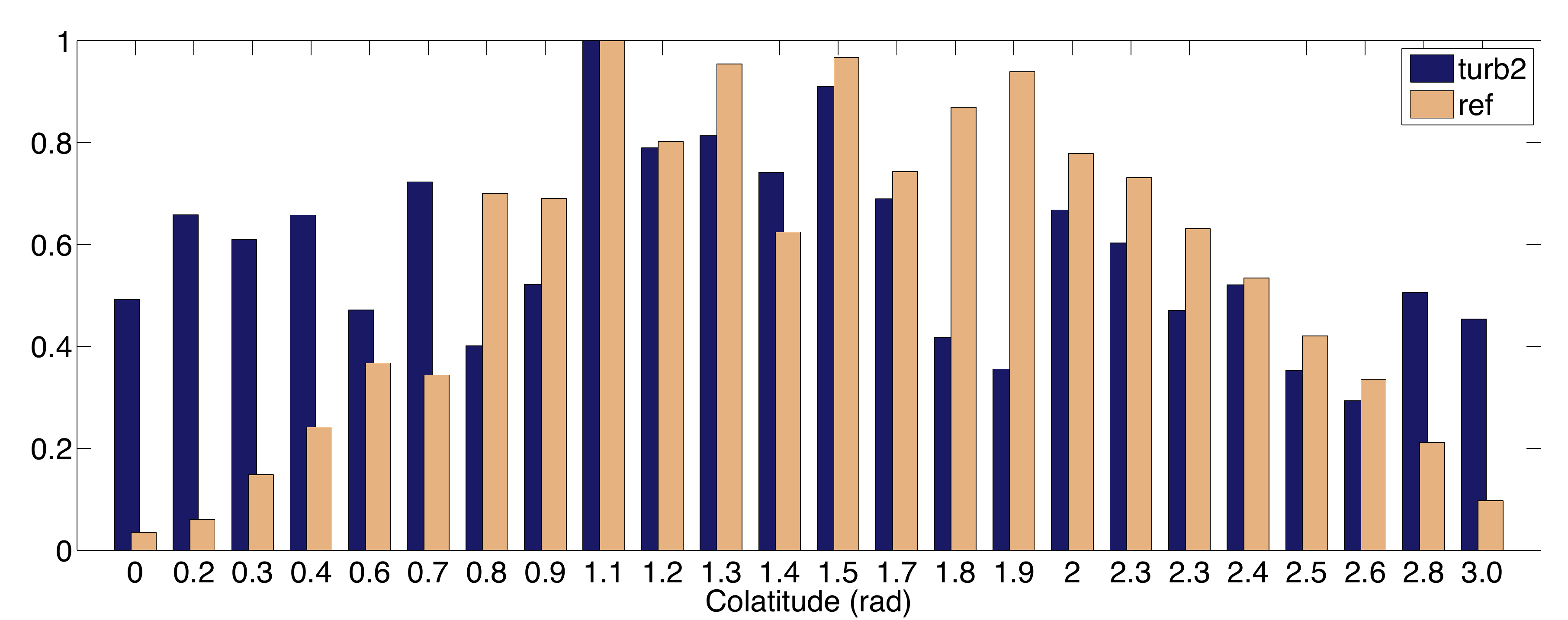}
  \caption{Distribution of the energy transported by the IGWs as a function of the latitude for two different models. The vertical axis is normalized
    by the maximum value.}
  \label{fig:e_lat}
\end{figure*}

We see that these values are indeed different for each plane and that they correspond to those expected by the theory. In particular, the $m=9$ peak is clearly
dominant in the $\theta_{9,9}$ plane and so is $m=0$ in the $\theta_{9,0}$ plane. That secondary peaks
are also visible can be explained by the following arguments. First, for the top lefthand panel ($\theta_{9,9}$), we notice two weak peaks at $m=12$ and $m=14$. We
assigned them to the mode $\ell=14,$ which is partially taken into account by the filter (see Fig. \ref{fig:zoom_plans}). Nevertheless, it is clear that
the $m=9$ mode is dominant. Then, the theory predicted that the existence of these planes assumes some approximation that is not verified
by our model. The adiabatic approximation made by \citet{goughHouches} is not true here since our waves are submitted to thermal and viscous
diffusivities. This could result in a leakage of the energy from one plane to another. Moreover, the presence of rotation in our model is to be 
considered as a small disturbance. Indeed, by using our 3D raytracing code (Mathis et al. 2015, in prep), we have shown that the propagation of IGWs in planes
is no longer verified in the case of rapidly rotating stars (except for $m=0$). Finally, we can invoke numerical noise because our
spherical harmonic transform uses a finite number of mesh points, which results in a leakage of the energy from one pair ($\ell,m$) to
another. {Despite these limitations, this is a clear example of the need to model those waves in full spherical geometry.}\\\newline
In \citet{Alvan:2014gx}, we showed that the energy of a given g mode tends to concentrate in high $m$ components. {We clarify this
  result in Fig. \ref{fig:e_lat} by showing the distribution of energy as a function of latitude for the current model (\textit{turb2}) and for a less
  turbulent model discussed in detail in that paper. Indeed, we see that for the model \textit{ref}, the energy is concentrated around
  the equator. We suspect that this concentration of energy may be due to the shape of convection (banana cells) in low-latitude regions. If we
  increase the turbulence of the convection (model \textit{turb2}), the distribution becomes more homogeneous in 
  latitude, but we still see a peak around the equator (colatitude $\theta=\pi/2$). } For instance, this could imply a more efficient transport of angular momentum induced
by the presence of IGWs in the equatorial regions.  

\section{Conclusion}
\label{sec:conclusion}

The aim of this paper has been to understand gravity waves in solar-like stars. To this end, we analyzed {our} 3D simulations in order to deepen
our understanding of the complexity of the spectrum excited. Here, we have explored and 
described the substructures of this spectrum by showing the co-existence of both progressive IGWs and standing modes. For the first time in 
simulation, we were able to visualize individual IGWs in the inner regions of the stars. 
\begin{itemize}
\item[$\bullet$] {Thanks to our frequency filtering method,
  we isolated some well-chosen waves in spectral space and shown that their energy paths in the real space correspond to the one
  predicted by the linear raytracing theory.} 
\item[$\bullet$] {We showed that it is possible to distinguish between g modes and progressive IGWs by measuring their
  damping rate. If the damping rate is too high, IGWs do not reach their inner turning point (whose depth depends on the frequency of the wave), and
  they cannot form resonant mode. In contrast, we see in the figures that isolated waves taken in the high part of the spectrum (high frequency
  and/or high degree $\ell$) can bounce at $r=r_{\rm{in}}$ and form a g mode. {The cut-off frequency separating g modes and progressive IGWs scales
    with $\left[\ell\left(\ell+1\right)\right]^{3/8}$}. {This analysis gives us precise knowledge of the composition of the
    spectrum excited in our model. Consequently, it gives a rather good understanding of the same processes acting on the real wave spectrum of the
    Sun.} }
\item[$\bullet$] {Our third result consists in the study of the spatial distribution of the waves energy in the 3D radiative zone. We have
    shown that, according to the theoretical predictions of \citet{goughHouches}, waves of different degrees $\ell$ and azimuthal order $m$ are
    distributed differently in space. Their energy is confined in planes, whose inclination with the polar direction varies with $(\ell,m)$. The plane
  that is the closest to the equator is the one with $m=\ell$. Moreover, as noted in \citet{Alvan:2014gx}, the waves associated to high values of $m$
   have the highest amplitude, which may be explained by studying the repartition of convective plumes. This result for the concentration of
  energy in planes is very important because it could guide our research into g modes at the surface of the Sun and our understanding of the
  angular momentum transport by gravity waves.}
\end{itemize}
{It is important to highlight that, to perform these simulations, we adopted values of diffusivities that are necessarily much higher than
  their microscopic values. For this reason, some of the predictions of the model (e.g., concerning the energetical aspects of the spectrum of IGWs
  excited by convection) must be considered with caution.}\\

The direct perspectives of this work are to improve our analysis by taking the effect of rotation that modifies the ray paths and break the propagation in
planes
into account. This project requires improving both the simulation (exploring other parameter regimes) and the theory. Indeed, our analysis is
guided by the results of the raytracing theory but also by theoretical results about the impact of rotation on the excitation and propagation of IGWs
\citep{BelkacemMathisetal2009,Mathis:2014hy}. {We also plan to study the nonlinear aspects of IGWs, including for example triadic interactions
  or wave breaking \citep[e.g.,][]{2002AnRFM..34..559S,Barker2011,2013JFM...723....1B}}. Finally, the extension of 
this study to other types of stars is in progress and will provide a new source of asteroseismic predictions.

\section*{Acknowledgments}
{We thank the referee, A. Barker, for his constructive comments that allowed us to improve the article. We are grateful to N. Featherstone,
  B. Brown, and D. Lecoanet for useful discussions. We acknowledge funding by the ERC grant STARS2 207430 
  (www.stars2.eu), {the European Communitys
    Seventh Framework Program ([FP7/2007-2013]) under the grant 
agreements no. 312844 (SPACEINN) and no. 269194 (IRSES/ASK),} and the CNRS Physique théorique et
  ses interfaces program. The simulations were performed using HPC resources of GENCI 1623 and PRACE
1069 projects.}

\section*{Appendix A: ASH code and 3D model description}
\label{sec:appendix-a}

\begin{figure}[h]
  \centering
  \includegraphics[width=0.45\textwidth]{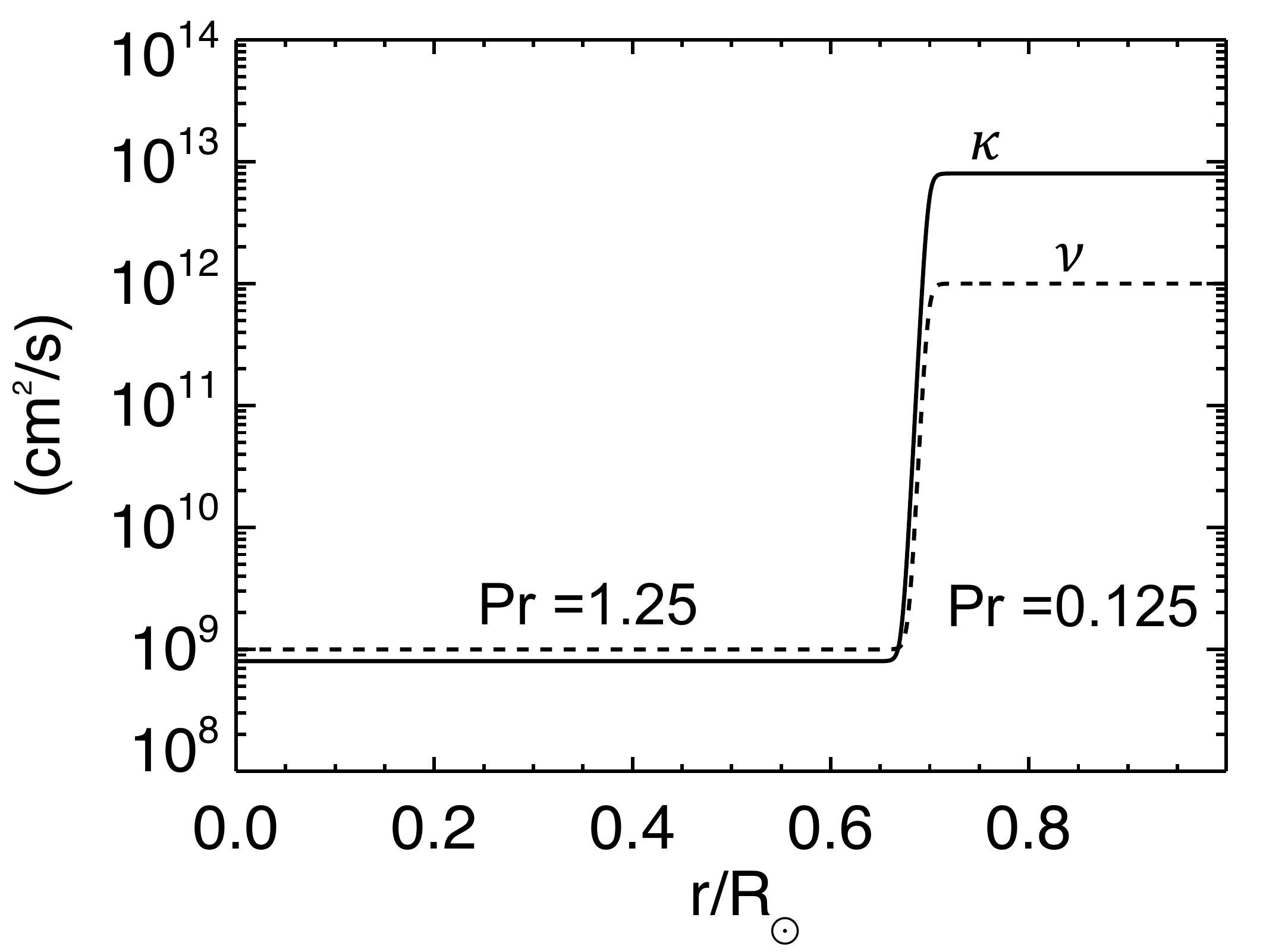}
  \caption{Radial profiles of the effective diffusivities $\kappa$ (thermal) and $\nu$ (viscous). The model used here is called \textit{turb2} in \citet{Alvan:2014gx}.}
  \label{fig:Knu}
\end{figure}

The ASH code solves the set of 3D anelastic hydrodynamic equations in a spherical geometry. The equations are fully nonlinear in velocity, and we
linearize the thermodynamic variables with respect to a spherically symmetric and evolving mean state. 
We denote $\bar\rho$, $\bar{{P}}$, $\bar {{T}}$, and $\bar {{S}}$
as the reference density, pressure, temperature, and specific
entropy. Fluctuations in this reference state are denoted by  $\rho$,
${P}$, ${T}$, and ${S}$. We assume a linearized equation of state
\begin{equation}
  \label{eq:1}
  \frac{\rho}{\bar\rho} = \frac{{P}}{\bar{{P}}} -
  \frac{{T}}{\bar{{T}}} = \frac{{P}}{\gamma\bar{{P}}} -\frac{{S}}{c_p}\hbox{,}
\end{equation}
{and the zeroth-order ideal gas law}
\begin{equation}
  \label{eq:2}
  \bar{{P}} = \mathcal{R} \bar\rho \bar{{T}} \hbox{,}
\end{equation}
where $\gamma$ is the first adiabatic exponent, $c_p$ the specific heat per
unit mass at constant pressure, and $\mathcal{R}$ the gas constant. \\\newline
We also introduce the local velocity ${\vec{\mathrm{v}}} = \left({\mathrm{v}}_r,{\mathrm{v}}_\theta,
  {\mathrm{v}}_\varphi \right)$ expressed in
spherical coordinates $\left( r,\theta,\varphi \right)$ in the frame rotating at constant angular velocity
$\vec\Omega_0=\Omega_0\vec{e_z}$. The set of hydrodynamic equations solved by ASH in the present case is
\begin{equation}
\label{eq:3}
  \left\{
      \begin{array}{ll}
\hbox{ (a) }&\vec\nabla.\left(\bar\rho\vec {{\mathrm{v}}}\right) = 0 \hbox{,}
\\
% \hbox{ (b) }&\bar\rho\left(\displaystyle\frac{\partial \vec {{\mathrm{v}}}}{\partial t} + \left(\vec
%     {{\mathrm{v}}}.\vec\nabla\right)\vec {{\mathrm{v}}}\right) =-\vec\nabla {P} + \rho\vec{{g}}-2\bar\rho \vec\Omega_0 \times \vec {{\mathrm{v}}} \\
% &\hbox{\hspace{1cm}}-
% \vec\nabla . \vec{\mathcal{D}} -\left[\vec\nabla\bar{{P}} - \bar\rho
%   \vec{{g}}\right] \hbox{,}\\
\hbox{ (b) }&\bar\rho\left(\displaystyle\frac{\partial \vec {{\mathrm{v}}}}{\partial t} + \left(\vec
    {{\mathrm{v}}}.\vec\nabla\right)\vec {{\mathrm{v}}}\right) = 
-\bar{\rho}\vec\nabla\widetilde\omega 
-\bar{\rho}\displaystyle\frac{S}{c_p}\vec{g} 
- 2\bar\rho \vec\Omega_0 \times \vec {{\mathrm{v}}} \\
&\hbox{\hspace{1cm}}- \vec\nabla . {\mathcal{D}} 
-\left[\vec\nabla\bar{{P}} - \bar\rho\vec{{g}}\right] \hbox{,}\\
\\
\hbox{ (c) }& \bar\rho \bar{{T}}\displaystyle\frac{\partial {S}}{\partial t} +
\bar\rho\bar{{T}}\vec{{\mathrm{v}}}.\vec\nabla\left({S}+\bar{{S}}\right)= \bar\rho \epsilon + \vec\nabla . \left[ \kappa_r\bar\rho c_p
  \vec\nabla\left({T}+\bar{{T}}\right) \right.\\
&\left. +\kappa\bar\rho\bar{{T}}\vec\nabla{S} +
  \kappa_0\bar\rho\bar{{T}}\vec\nabla\bar{{S}} \right] +2\bar\rho\nu\left[e_{ij}e_{ij}-1/3\left(\vec\nabla
    . \vec{{\mathrm{v}}}\right)^2\right] \hbox{,}
      \end{array}
  \right.
\end{equation}
where equation (a) is the continuity equation in the anelastic approximation, (b) the momentum equation, and (c)  the energy equation.\\
We define the gravitational acceleration $\vec{{g}}$, the viscous stress tensor $\mathcal{D}$ defined by
\begin{equation}
  \mathcal{D}_{ij} = -2\bar\rho\nu\left(e_{ij} - 1/3\left(\vec\nabla  . \vec{{\mathrm{v}}}\right)\delta_{ij}\right)\hbox{,}
\end{equation}
with $e_{ij}=1/2\left( \partial_j {\mathrm{v}}_i+\partial_i {\mathrm{v}}_j  \right)$ the strain rate tensor and $\delta_{ij}$ the Kronecker
symbol. The bracketed term on the righthand side of the momentum equation is initially zero since the system
begins in hydrostatic balance. For this equation, we used the Lantz-Braginsky-Roberts (LBR) form \citep[e.g.,][]{LantzPHD,1995GApFD..79....1B} advocated by \citet{Brown:2012bd},
who have shown that the classical anelastic formulation may introduce a bias in the conservation of energy, in the case of strongly stably stratified
atmospheres. In this formulation, we use the reduced pressure $\widetilde{\omega}=P/\bar\rho$ instead of the fluctuating pressure $P$ and neglect
the buoyancy term $\bar\rho\widetilde{\omega}\vec\nabla\left(\displaystyle{\bar{S}}/{c_P}\right)$ associated with the
non-adiabatic reference state in the radiative zone. Thus, the new buoyancy term $\bar{\rho}\vec{g}\displaystyle{S}/{c_p}$ only contains the
contribution of entropy fluctuations, and the contribution due to pressure perturbations is included in the reduced pressure gradient. In the energy
equation (c), $\kappa_r$ is the radiative diffusivity based on a 1D solar structure model, and $\nu$ and $\kappa$ are effective diffusivities
modeling momentum and heat transport by subgrid-scale (SGS) motions that are unresolved by the simulation. Their profiles are represented in
Fig. \ref{fig:Knu}.

The diffusivity $\kappa_0$ is also part of the SGS treatment in the convective zone and is chosen such that the entropy
flux carries the solar flux outward at the top of the 
domain. It is negligible in the radiative zone \citep{2000ApJ...532..593M}. Finally, the volume heating term $\bar\rho \epsilon$ represents the
energy generation by nuclear burning with $\epsilon = \epsilon_0 {T}^k$. The constant $\epsilon_0$ is set such that the radially integrated heating
term equals the solar luminosity at the base of the convection zone, and $k = 9$ allows us to take both contributions of the p-p chains
and CNO cycles into account \citep{BrunAl2002}. \\\newline
The computational domain extends from $r_{\mathrm{bot}}$ = 0 to $r_{\mathrm{top}}$ = 0.97$R_\odot$ where
$R_\odot =6.9599 \times10^{10}$ cm is the solar radius. The boundary conditions at the top of the domain are torque-free 
velocity conditions and constant heat flux \citep{Brun:2011bl}. The inner boundary conditions are chosen to deal with the central singularity at
$r_{\mathrm{bot}}$ = 0. Conditions on the poloidal and toroidal components of the mass flux impose that only $\ell=1$ mode can go through the center,
and thermal condition are also adapted. This treatment is explained in detail in \citet{Alvan:2014gx}. \\
In the model presented here, the spatial resolution is $N_r \times N_\theta \times N_\varphi = 1581 \times 512 \times 1024$, with a non-uniform radial
grid chosen to resolve the turbulent convection, the oscillations of IGWs in the radiative zone, and the steep drop of diffusivities at the interface between
both zones \citep{Alvan:2014gx}.\\\newline
The 3D simulation is initialized using a reference state derived from a 1D solar structure model \citep{BrunAl2002}. The set of equations
\eqref{eq:3} is then solved in expanding the velocity and thermodynamic variables in spherical harmonics $Y_{\ell,m}(\theta,\varphi)$ for their 
horizontal structure and using a fourth-order finite-difference approach on a nonuniform grid for their radial part. For the time integration, we use an explicit
Adams-Bashforth scheme for the advection and Coriolis terms and a semi-implicit Crank-Nicolson treatment for the diffusive and buoyancy terms
\citep{1984JCoPh..55..461G,CluneAl1999}. The reference state is updated during the
simulation with the spherically averaged perturbation fields. \\\newline 
After a few convective overturning times, a balance {is established} between the
contributions {to the total flux} of the different {physical} processes \citep{Brun:2011bl,Alvan:2014gx}. It allows us to see that the convective flux becomes negative
at the base of the convective zone, which is the signature of the overshoot process \citep[e.g.,][]{1991AA...252..179Z}, where downflows penetrate the radiative zone slightly owing to their inertia. {This induces a perturbation in velocity and temperature at the top of the
  radiative zone, which propagates in the form of a gravity wave. With the bulk excitation process \citep{Lecoanet:ws}, overshoot is responsible for
  the excitation of IGWs in solar-like stars.}

\section*{Appendix B: Understanding the cut-off frequency between modes and progressive waves}

The goal of this appendix is to derive the dependence of the cut-off frequency that separates modes and progressive waves, $\omega_{\rm c}$, on their latitudinal degree ($\ell$). 

We develop the radial component of the velocity on spherical harmonics:
\begin{equation}
u_{r}=\sum_{\ell,m}u_{r;\ell,m}\left(r\right)Y_{l,m}\left(\theta,\varphi\right)\exp\left(i\omega t\right).
\end{equation} 
Next, we get from the system formed by the linearized momentum, continuity, and heat transport equations in the adiabatic case
\begin{equation}
\frac{{\rm d}^2\Psi_{\ell,m}\left(r\right)}{{\rm d}r^2}+k_{V;\ell}^{2}\left(r\right)\Psi_{\ell,m}\left(r\right)=0,
\end{equation}
where$\quad\Psi_{\ell,m}={\overline\rho}^{\frac{1}{2}}r^2 u_{r;\ell,m}$, and
\begin{equation}
k_{V;\ell}^{2}=\left(\frac{N^2}{\omega^2}-1\right)\frac{\ell\left(\ell+1\right)}{r^2}.
\end{equation}
From then on, we focus on the low-frequency asymptotic regime in which $\omega\ll N$. We follow \cite{ZahnTalonMatias1997} by applying the JWKB and the quasi-adiabatic approximations that allows us to derive the expression for $u_{r;\ell,m}$:
\begin{eqnarray}
u_{r;\ell,m}&=&A_{\ell,m}r^{-\frac{3}{2}}{\overline\rho}^{-\frac{1}{2}}\left(\frac{N^2}{\omega^2}\right)^{-\frac{1}{4}}P_{l}^{\vert m\vert}\left(\cos\theta\right)\nonumber\\
&&\times\cos\left(\omega t+m\varphi\pm\int_{r}^{{\hat r}_{\rm out}}k_{V;\ell}{\rm d}r'\right)\exp\left[-\frac{\tau\left(r,\ell,\omega\right)}{2}\right],
\label{eq:urlm}
\end{eqnarray} 
where $A_{\ell,m}$ is an amplitude coefficient, which includes the normalization of spherical harmonics, and $P_{\ell}^{m}$ are the associated Legendre polynomials. We recall the radiative damping expression (Eq.~\ref{eq:tau})
\begin{equation}
  \tau\left(r,\ell,\omega\right) =
  \left[\ell(\ell+1)\right]^{\frac{3}{2}}\int_{r}^{{\hat r}_{\rm
      out}}\kappa\frac{N^3}{\omega^4}\frac{\mathrm dr'}{r'^3}.
\end{equation}
Since the JWKB approximation falls at the two turning points $\left(r_{\rm in},r_{\rm out}\right)$, we introduce the two radius $\left({\hat r}_{\rm in},{\hat r}_{\rm out}\right)$ (with ${r}_{\rm in}<{\hat r}_{\rm in}$ and ${\hat r}_{\rm out}<{r}_{\rm out}$) between which it can be applied (see the detailed discussion in \cite{BM1972} and \cite{2013AA...553A..86A}).\\ 

Then, we derive the horizontally averaged kinetic energy of IGWs using results derived by \cite{ZahnTalonMatias1997}:
\begin{equation}
E_{\rm K;\ell,m}\left(r\right)=\frac{1}{2}{\overline\rho}\left<{\vec u}_{\ell,m}^2\right>_{\theta,\varphi}=\frac{1}{2}\frac{N^2}{\omega^2}{\overline\rho}\left<{u_{r;\ell,m}^{2}}\right>_{\theta,\varphi},
\end{equation}
where $\left<\cdot\!\cdot\!\cdot\right>_{\theta,\varphi}=1/4\pi\int_{\theta,\varphi}\sin\theta{\rm d}\theta{\rm d}\varphi$. Using Eq. (\ref{eq:urlm}), it becomes
\begin{equation}
E_{\rm K;\ell,m}\left(r\right)=\frac{1}{4}A_{\ell,m}^2\frac{N}{\omega}\left<\left[P_{l}^{\vert m\vert}\left(\cos\theta\right)\right]^2\right>_{\theta}r^{-3}\exp\left[-{\tau\left(r,\ell,\omega\right)}\right].
\label{eq:kinE}
\end{equation}
We then define the cut-off frequency that separates modes and progressive waves with the following criterion:
\begin{equation}
E_{\rm K;\ell,m}\left({\hat r}_{\rm in}\right)=K E_{\rm K;\ell,m}\left({\hat r}_{\rm out}\right).
\label{eq:criterion}
\end{equation}
When $K\ll1$, the wave has lost a large part of its kinetic energy and cannot reflect to form a standing mode. Using Eq. (\ref{eq:kinE}), it becomes
\begin{equation}
\exp\left[-\tau\left({\hat r}_{\rm in},\ell,\omega\right)\right]=K\frac{N\left({\hat r}_{\rm out}\right){\hat r}_{\rm out}^{-3}}{N\left({\hat r}_{\rm in}\right){\hat r}_{\rm in}^{-3}}=\alpha. 
\end{equation}
Assuming that ${\hat r}_{\rm in}$ and ${\hat r}_{\rm out}$ vary weakly with $\omega_{\rm c}$ for $\omega_{\rm c}\ll N$, we finally obtain
\begin{equation}
\omega_{\rm c}\approx\left[-\frac{\displaystyle\int_{{\hat r}_{\rm in}}^{{\hat r}_{\rm out}}\kappa N^3\frac{\mathrm dr'}{r'^3}}{\ln\alpha} \left[\ell(\ell+1)\right]^{\frac{3}{2}}\right]^{\frac{1}{4}}\equiv\beta\left[\ell\left(\ell+1\right)\right]^{\frac{3}{8}} 
\end{equation}
as observed in power spectrum of our direct 3D nonlinear ASH simulations (Fig. \ref{fig:Bigfig}).
%=========%=========%=========%=========%=========%=========%=========%=
%BIBLIOGRAPHIE
%=========%=========%=========%=========%=========%=========%=========%=
\bibliographystyle{aa}  % A&A bibliography style file (aa.bst)
\bibliography{Library} % your references in file: Yourfile.bib

%=========%=========%=========%=========%=========%=========%=========%=
%THE END
%=========%=========%=========%=========%=========%=========%=========%=
 \end{document}